\documentstyle[aps,preprint,prb,eqsecnum,tighten,epsf]{revtex}
\preprint{\underline{cond--mat/9701159, submitted to Phys. Rev. B}}
\begin{document}
\draft
\title{Theory of a spherical quantum rotors model: \\
low--temperature regime and finite--size scaling}
\date{November 26, 1996}
\author{Hassan Chamati$^1$, Ekaterina S. Pisanova$^2$ and
Nicholay S. Tonchev$^1$\cite{nstonchev}}
\address{$^1$Institute of Solid State Physics, Bulgarian Academy of
Sciences, \\Tzarigradsko chauss\'ee 72,
1784 Sofia, Bulgaria}
\address{$^2$University of Plovdiv, Faculty of Physics, 24 "Tzar Assen"
str.,\\ 4000 Plovdiv, Bulgaria}
\maketitle
\begin{abstract}
The quantum rotors model can be regarded as an effective model for the
low--temperature behavior of the quantum Heisenberg antiferromagnets.
Here, we consider a $d$--dimensional model in the spherical
approximation confined to a general geometry of the form
$L^{d-d'}\!\!\times\!\!\infty^{d'}\!\!\times\!\! L_{\tau}^{z}$ (
$L$--linear space size and $L_{\tau}$--temporal size) and subjected to
periodic boundary conditions. Due to the remarkable opportunity it
offers for rigorous study of finite--size effects at arbitrary
dimensionality this model may play the same role in quantum critical
phenomena as the popular Berlin--Kac spherical model in classical
critical phenomena. Close to the zero--temperature quantum
critical point, the ideas of finite--size scaling are utilized to the
fullest extent for studying the critical behavior of the model. For
different dimensions $1<d<3$ and $0\leq d'\leq d$ a detailed analysis,
in terms of the special functions of classical mathematics, for the
susceptibility and the equation of state is given. Particular
attention is paid to the two--dimensional case.
\end{abstract}
\narrowtext
\section{Introduction}\label{Intr}
In recent years there has been a renewed
interest~\cite{Belitz94,sachdev96,Sondhi96}
in the theory of zero--temperature quantum phase transitions initiated
in 1976 by Hertz's quantum dynamic renormalization group
\cite{Hertz76} for itinerant ferromagnets. Distinctively from
temperature driven critical phenomena, these phase transitions occur
at zero temperature as a function of some non--thermal control
parameter (or a competition between different parameters describing
the basic interaction of the system), and the relevant fluctuations
are of quantum rather than thermal nature.

It is well known from the theory of critical phenomena that for
the temperature driven phase transitions quantum effects are
unimportant near critical points with $T_{c}>0$. It could be
expected, however, that at rather small (as compared to
characteristic excitation in the system) temperature, the leading
$T$ dependence of all observables is specified by the properties of
the zero--temperature critical points, which take place in
quantum systems. The dimensional crossover rule asserts that the
critical singularities of such a quantum system at $T=0$ with
dimensionality $d$ are formally equivalent to those of a classical
system with dimensionality $d+z$ ($z$ is the dynamical critical
exponent) and critical temperature $T_{c}>0$. This make it possible to
investigate low--temperature effects (considering an effective system
with $d$ infinite space and $z$ finite time dimensions) in the
framework of the theory of finite--size scaling (FSS). The idea of
this theory has been applied to explore the low--temperature
regime in quantum systems (see Refs.~\onlinecite{chakravarty89,%
chubukov94,sachdev94}),
when the properties of the thermodynamic observables in the {\it
finite--temperature quantum critical region} have been the main focus
of interest. The very {\it quantum critical region} was introduced
and studied first by Chakravarty et al~\cite{chakravarty89} using
the renormalization group methods. The most famous model for
discussing these properties is the quantum nonlinear ${\cal O}(n)$
sigma model (QNL$\sigma$M).\cite{chakravarty89,chubukov94,sachdev94,%
Rosenstein90,Castro93,fujii95}

Recently an equivalence between the QNL$\sigma$M
in the limit $n\rightarrow\infty$ and a quantum version of the
spherical model or more precisely the "spherical quantum rotors" model
(SQRM) was announced.\cite{vojta95} The SQRM is an interesting model
in its own. Due to the remarkable opportunity it offers for rigorous
study of finite--size effects at arbitrary dimensionality SQRM may
play the same role in quantum critical
phenomena as the popular Berlin--Kac spherical model in classical
critical phenomena. The last one became a touchstone for various
scaling hypotheses and source of new ideas in the general theory of
finite--size scaling (see for example Refs.~\onlinecite{singh85,%
brankov88,privman90,singh92,brankov94,allen94,allen95,chamati96}
and references therein). Let us note that an increasing interest
related with the spherical approximation (or large  $n$--limit)
generating tractable models in quantum critical phenomena has been
observed in the last few years.\cite{vojta95,tu94,nieu95,vojta96}

In Ref.~\onlinecite{vojta95}, the critical exponents for the
zero--temperature quantum fixed point and the finite--temperature
classical one as a function of dimensionality was obtained. What
remains beyond the scope of Ref.~\onlinecite{vojta95} is to study
in an exact manner the scaling properties of the model in different
regions of the phase diagram including the {\it quantum critical
region} as a function of the dimensionality of the system. In the
context of the finite--size scaling theory both cases: ({\bbox i})
The infinite $d$--dimensional quantum system at low--temperatures
$\infty^{d}\times\!\! L_{\tau}^{z}$
($L_{\tau}\sim\left(\frac{\hbar}{k_{B}^{}T}\right)^{1/z}$ is the
finite--size in the imaginary time direction) and ({\bbox{ii}}) the
finite system confined to the geometry
$L^{d-d'}\!\!\times\!\!\infty^{d'}\!\!\times\!\! L_{\tau}^{z}$
($L$--linear space size) are of crucial interest.

Earlier a class of exactly solvable lattice models intended to
study the displacive structural phase transition have been
intensively considered in both finite--size and bulk geometry.%
\cite{verbeure92,pisanova93,chamati94,pisanova95}
The main feature of these models is that the real anharmonic
interaction is substituted with its quantum mean spherical
approximation reducing the problem to an exactly solvable one.
We expect that the analytical technique proposed below will
apply to these models too.

In this paper a detailed theory of the scaling properties of the
SQRM with nearest-neighbor interaction is presented. The plan of the
paper is as follows: we start with a brief review of the model and
the basic equation for the quantum spherical field in the case of
periodic boundary conditions (Section~\ref{model}). Since we would
like to exploit the ideas of the FSS theory, the bulk system in the
low--temperature region is considered like an effective
($d+1$)--dimensional classical system with one finite (temporal)
dimension. This is done to enable contact to be made with other
results based on the spherical type approximation e.g. in the
framework of the spherical model and the QNL$\sigma$M in the limit
$n\rightarrow\infty$ (Section~\ref{bulk}). In Section~\ref{fing} we
consider FSS form of the spherical field equation for the system
confined to the general geometry
$L^{d-d'}\!\!\times\!\!\infty^{d'}\!\!\times\!\! L_{\tau}^{z}$. This
equation turns out to allow for analytic studies of the finite--size
and low--temperature asymptotes for different $d$ and $d'$. Special
attention is laid on the two dimensional system. The remainder of the
paper contains the details of the calculations:
Appendices~\ref{app1},~\ref{app2}.

\section{The Model}\label{model}
The model we will consider here describes a magnetic
ordering due to the interaction of quantum spins. This has
the following form\cite{vojta95}
\begin{equation}
{\cal H} = \frac{1}{2} g \sum_{\ell} {\cal P}_{\ell}^{2} -
\frac{1}{2} \sum_{\ell\ell'} {\bbox J}_{\ell\ell'}^{} {\cal S}_{\ell}^{}
{\cal S}_{\ell'}^{}+ \frac{\mu}{2} \sum_{\ell} {\cal
S}_{\ell}^{2}  - H \sum_{\ell} {\cal S}_{\ell}^{}, \label{model1}
\end{equation}
where ${\cal S}_{\ell}^{}$ are spin operators at site $\ell$, the
operators ${\cal P}_{\ell}^{}$ are "conjugated" momenta (i.e.
$[{\cal S}_{\ell}^{},{\cal S}_{\ell'}^{}] = 0$, $[{\cal P}_{\ell}^{},
{\cal P}_{\ell'}^{}]
= 0$, and $[{\cal P}_{\ell}^{},{\cal S}_{\ell'}^{}] =
i\delta_{\ell\ell'}^{}$, with $\hbar =1$), the coupling constants
${\bbox J}_{\ell,\ell'}^{}={\bbox J}$ is between nearest neighbors
only,~\cite{dyn} the coupling
constant $g$ is introduced so as to measure the strength of the
quantum fluctuations (below it will be called quantum parameter),
$H$ is an ordering magnetic field, and finally the spherical field
$\mu$ is introduced so as to insure the constraint
\begin{equation}
\sum_{\ell} \left<{\cal S}_{\ell}^{2}\right> = N. \label{model2}
\end{equation}
Here $N$ is the total number of the quantum spins located at
sites "$\ell$" of a hypercubical lattice of size
$L_{1}\times L_{2}\times\cdots\times L_{d}=N$ and
$\left<\cdots\right>$ denotes the standard thermodynamic average
taken with ${\cal H}$.

Let us note that the commutation relations for the operators
${\cal S}_{\ell}^{}$ and ${\cal P}_{\ell}^{}$ together with the
quadratic kinetic term in the Hamiltonian~(\ref{model1}) do not
describe quantum Heisenberg--Dirac spins but quantum rotors as it
was pointed out in Ref.~\onlinecite{vojta95}.

Under periodic boundary conditions, Eq.~(\ref{model2}) takes the form
\begin{equation}\label{model3}
1=\frac{\lambda}{2N} \sum_{\bbox q} \frac{1}{\sqrt{\phi+2\sum_{i=1}^{d}
(1-\cos q_{i})}} \coth\left(\frac{\lambda}{2t}\sqrt{\phi +
2\sum_{i=1}^{d} (1-\cos q_{i})} \right) + \frac{h^{2}}{\phi^{2}},
\end{equation}
where we have introduced the notations: $\lambda=\sqrt{\frac{g}
{\bbox J}}$ is the normalized quantum parameter,  $t=\frac{T}{\bbox
J}$--the normalized temperature, $h=\frac{H}{\sqrt{\bbox J}}$ -- the
normalized magnetic field, $b = \frac{2\pi t}{\lambda}$, and $\phi
= \frac{\mu}{\bbox J} -2d$ is the shifted spherical field.

In Eq.~(\ref{model3}) the vector $\bbox q$ is a collective symbol,
which has for $L_{j}$ odd integers the components:
$$
\left\{\frac{2\pi n_{1}}{L_{1}},\cdots,\frac{2\pi
n_{d}}{L_{d}}\right\}, \ \ n_{j}
\in\left\{-\frac{L_{j}-1}{2},\cdots,\frac{L_{j}-1}{2}\right\}.
$$

A previous direct analysis~\cite{vojta95} of Eq.~(\ref{model3}) in
the thermodynamic limit shows that there can be no long--range
order at finite temperature, for $d\leq2$ (in accordance with the
Mermin--Wagner theorem). For $d>2$ one can find long--range
order at finite temperature up to a critical temperature
$t_{c}(\lambda)$. Here we shall consider the low--temperature
region for $1<d<3$.
\section{The infinite system}\label{bulk}

In the thermodynamic limit the $d$-dimensional sum over the
momentum vector $\bbox q$ in Eq.~(\ref{model3}) changes in $d$
integrals over the $q_{i}$'s in the first Brillouin zone and the
equation for the shifted spherical field $\phi$ reads
\begin{equation}\label{bulk1}
1=\frac{t}{(2\pi)^{d}} \sum_{m=-\infty}^{\infty}
\int_{-\pi}^{\pi}dq_{1}\cdots\int_{-\pi}^{\pi}dq_{d}
\frac{1}{\phi+2\sum_{i=1}^{d} (1-\cos q_{i}) +b^{2} m^{2}} +
\frac{h^{2}}{\phi^{2}}.
\end{equation}

After some algebra (see Appendix~\ref{app1}), Eq.~(\ref{bulk1})
takes the form ($1<d<3$)
\begin{equation}\label{bulk7}
\frac{1}{\lambda}-\frac{1}{\lambda_{c}}=-\frac{1}{(4\pi)^{(d+1)/2}}
\left|\Gamma \left( \frac{1-d}{2} \right) \right|\phi^{(d-1)/2} +
\frac{2}{(4\pi)^{(d+1)/2}}\phi^{(d-1)/2} {\cal K}
\left(\frac{d-1}{2},\frac{\lambda}{2t}\phi^{1/2}\right) +
\frac{h^{2}}{\phi^{2}},
\end{equation}
where $\lambda_{c}$ is the quantum critical point and
\begin{equation}
{\cal K}(\nu,y)\equiv {\cal K}_{1}\left.\left(\frac{d-1}{2}
\right|1,y\right)=2 \sum_{m=1}^{\infty} (y m)^{-\nu}K_{\nu}(2my).
\end{equation}
Here $K_{\nu}(x)$ is the MacDonald function (second modified Bessel
function). The asymptotic forms of the functions ${\cal K}_{1}\left.
\left(\frac{d-1}{2}\right|1,y\right)$ are studied in
Appendix~\ref{app2}. In the remainder of this section we will study
the effect of the temperature on the susceptibility and the equation
of state near the quantum critical fixed point.

\subsection{Zero--field susceptibility}
\begin{figure}
\epsfxsize=4.5in
\centerline{\epsffile{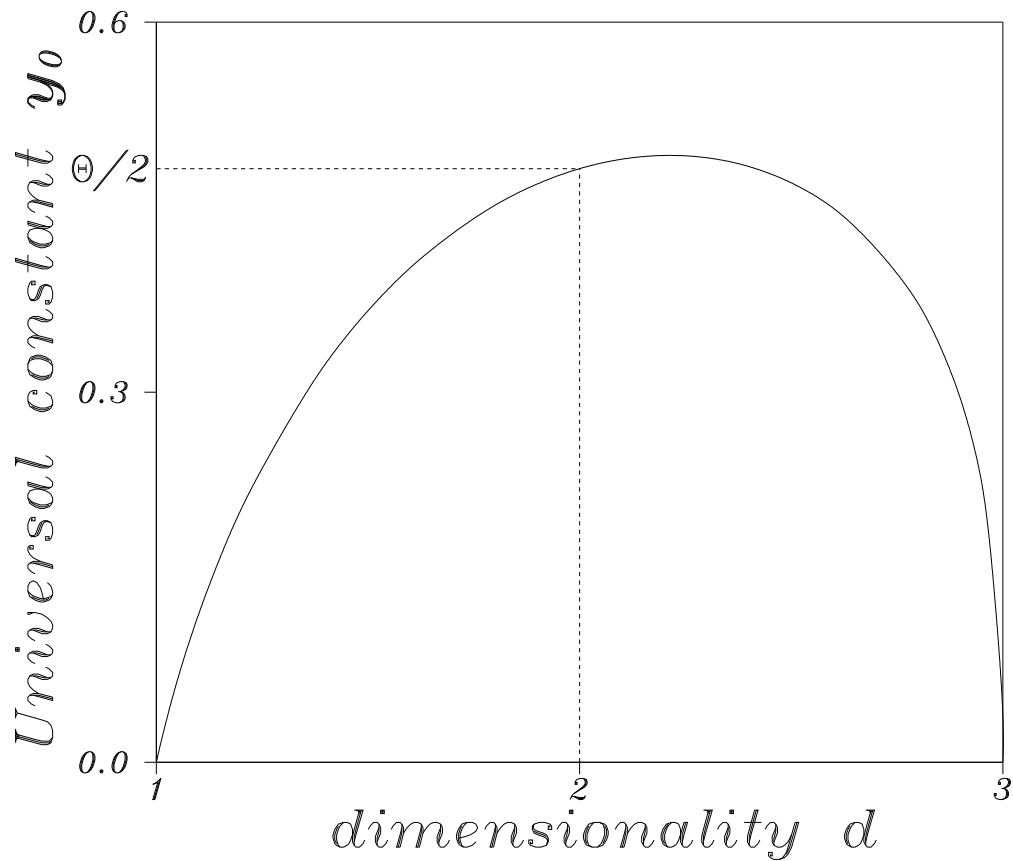}}
\vspace{0.1in}
\caption{The dependence of the universal constant $y_{0}$
upon the dimensionality $d$. The constant $\Theta=0.962424...$
is obtained for the two--dimensional system (see
Eq.~(\protect\ref{theta}))}\label{fig1}
\end{figure}

After making vanishes the field $h$, from Eq.~(\ref{bulk7}) we find
that the normalized zero--field susceptibility $\chi=\phi^{-1}$
on the line $\lambda = \lambda_{c}$ ($t\rightarrow0^{+}$) is given by
\begin{equation}\label{bulk8}
\chi = \frac{\lambda_{c}^{2}}{4 y_{0}^{2}} t^{-2},
\end{equation}
where $y_{0}$ is the universal solution of
\begin{equation}\label{scal1}
\left|\Gamma \left( \frac{1-d}{2} \right) \right| =
2 {\cal K}\left(\frac{d-1}{2},y\right).
\end{equation}
The behavior of the universal constant $y_{0}$ as a function of the
dimensionality $d$ of the system is shown in FIG.~\ref{fig1}.

One can see that the low--temperature behavior of
the susceptibility increases as the inverse
of the square of the temperature above the quantum critical point.

In what follows we will try to investigate Eq.~(\ref{bulk7}) for
different dimensions ($1<d<3$) of the system and in different regions
of the ($t,\lambda$) phase diagram.
Introducing the "shifted" critical value of the quantum parameter
due to the temperature by
\begin{equation}\label{bulk10}
\frac{1}{\lambda_{c}^{\mp}(t)}\approx\frac{1}{\lambda_{c}}
\mp\frac{1}{2\pi^{(d+1)/2}}\left(\frac{t}{\lambda_{c}}\right)^{d-1}
\Gamma\left(\frac{d-1}{2}\right)|\zeta(d-1)|,
\end{equation}
(where $\zeta(x)$ is the Rieman zeta function)
one has to make a difference between the two cases $d<2$ "sign
$-$" and $d>2$ "sign $+$". In the first case ($1<d<2$), it is possible
to define the {\it quantum critical region} by the inequality
\begin{equation}
\left|\frac{1}{\lambda}-\frac{1}{\lambda_{c}}\right|\ll
\frac{1}{2\pi^{(d+1)/2}}\left(\frac{t}{\lambda_{c}(t)}\right)^{d-1}
\Gamma\left(\frac{d-1}{2}\right)|\zeta(d-1)|.
\end{equation}

For $1<d<2$ the function ${\cal K}(\nu,y) \sim y^{-1}$ and
by substitution in~(\ref{bulk7}) we obtain for $\lambda<\lambda_{c}$
(outside of the {\it quantum critical region})
\begin{equation}\label{suscepb}
\chi \approx
\left[\frac{\left|\Gamma\left(1-\frac{d}{2}\right)\right|}{(4\pi)
^{d/2}\delta\lambda}
\right]^{2/(d-2)} t^{2/(d-2)},
\end{equation}
where
$$
\delta\lambda=\frac{1}{\lambda_{c}}-\frac{1}{\lambda}.
$$
In Eq.~(\ref{suscepb}) we see that the susceptibility is going to infinity
with power law degree when the quantum fluctuations become important
($t\rightarrow0^{+}$) and there is no phase transition driven
by $\lambda$ in the system for dimensions between 1 and 2.

In the second case ($2<d<3$), one has
\begin{equation}\label{bulk11}
\chi \approx
\left[\frac{\left|\Gamma\left(1-\frac{d}{2}\right)\right|}{(4\pi)
^{d/2}}\frac{\lambda_{c}(t)}{\lambda-\lambda_{c}(t)}
\right]^{2/(d-2)}
t^{2/(d-2)},
\end{equation}
as a solution for $\lambda$ less than $\lambda_{c}$ and greater than
the critical value $\lambda_{c}(t)$ of the quantum parameter. Here for
finite temperatures there is a phase transition driven by the quantum
parameter $\lambda$ with critical exponent of the $d$--dimensional
classical spherical model $\gamma= \frac{2}{d-2}$. This however is
valid only for very close values of $\lambda$ to $\lambda(t)$. For
$\lambda<\lambda_{c}(t)$ the susceptibility is infinite.

In the region where $\lambda>\lambda_{c}$ the zero--field
susceptibility is given by
\begin{equation}\label{bulk12}
\chi\approx\left[\frac{(4\pi)^{(d+1)/2}}{\Gamma\left(\frac{1-d}{2}
\right)}\delta\lambda
\right]^{\frac{2}{1-d}}.
\end{equation}
This results is valid for every $d$ between the lower and the upper
quantum critical dimensions i.e. $1<d<3$.

The important case $d=2$
can be solved easily and one gets
\begin{equation}\label{bulk13}
\phi^{1/2}=\frac{2t}{\lambda}\text{arcsinh}\left\{\frac{1}{2}
\exp\left[\frac{2\pi\lambda}{t}\delta\lambda
\right]\right\}
\end{equation}

For the susceptibility, Eq.~(\ref{bulk13}) yields
\begin{mathletters}\label{bulk14}
\begin{equation}
\chi \approx \frac{\lambda^{2}}{t^{2}} \exp\left(-\frac{4\pi\lambda_{c}}{t}
\delta\lambda
\right)
\end{equation}
for $\frac{2\pi}{t}\left|\frac{\lambda}{\lambda_{c}}-1\right|\gg1$ and
$\lambda < \lambda_{c}$ i.e. in the renormalized classical region.
For $\lambda=\lambda_{c}=3.1114...$
\begin{equation}
\chi=\frac{1}{\Theta^{2}}\left(\frac{\lambda_{c}}{t}\right)^{2},
\end{equation}
where the universal constant
\begin{equation}\label{theta}
\Theta= 2y_{0}=2\ln\left(\frac{\sqrt{5}+1}{2}\right)=
-2\ln\left(\frac{\sqrt{5}-1}{2}\right)=.962424...
\end{equation}
was first obtained in the framework of the 3-dimensional classical
mean spherical model with one finite dimension.\cite{singh85}
Finally for
$\frac{2\pi}{t}\left|\frac{\lambda}{\lambda_{c}}-1\right|\gg1$ and
$\lambda > \lambda_{c}$ i.e. in the quantum disordered region.
\begin{equation}\label{bulk14a}
\chi\approx\left[4\pi\delta\lambda
\right]^{-2}\left\{1+\frac{2t}{\pi\lambda_{c}\delta\lambda}
\exp\left[\frac{4\pi\lambda_{c}}{t}\delta\lambda
\right]\right\}.
\end{equation}
\end{mathletters}
The first term of Eq.~(\ref{bulk14a}) is a particular case of
Eq.~(\ref{bulk12}) for $d=2$.

From Eqs.~(\ref{bulk14}) one can transparently see the
different behaviors of $\chi(T)$ in tree regions: a) renormalized
classical region with exponentially divergence as $T\rightarrow 0$,
b) {\it quantum critical region} with $\chi(T)\sim T^{-2}$ and
crossover lines $T\sim|\lambda-\lambda_{c}|$, and c) quantum
disordered region with temperature independent susceptibility (up to
exponentially small corrections) as
$T\rightarrow 0$. The above results~(\ref{bulk13}) and~(\ref{bulk14})
coincide in form with those obtained in
Refs.~\onlinecite{chubukov94,Rosenstein90} for the two dimensional
QNL$\sigma$M in the $n\rightarrow\infty$ limit. The only differences
are: in Eq.~(\ref{bulk13}) the temperature is scaled by $\lambda$, and
the critical value $\lambda_{c}$ is given by Eq.~(\ref{bulk6}), while
for the QNL$\sigma$M it depends upon the regularization scheme.

\subsection{Equation of state}
The equation of state of the model hamiltonian~(\ref{model1}) near the
quantum critical point is obtained after substituting the shifted
spherical field $\phi$ by the magnetization $\cal M$ through the
relation
\begin{equation}\label{mag}
{\cal M} =\frac{h}{\phi},
\end{equation}
in Eq.~(\ref{bulk7}), which allows us to write the equation of state
in a scaling form
\begin{equation}\label{bulk15}
-\frac{\delta\lambda}{{\cal M}^{1/\beta}} + (4\pi)^{-(d+1)/2}
\left[\frac{h}{{\cal M}^{\delta}}\right]^{1/\gamma}\left\{\left|
\Gamma\left(\frac{1-d}{2}\right)\right|- 2{\cal K}\left(\frac{d-1}{2},
\frac{\lambda}{2}\left(\frac{{\cal M}^{\nu/\beta}}{t}
\right)\left(\frac{h}{{\cal M}^{\delta}}\right)^{\beta}\right)
\right\}=1.
\end{equation}
We conclude that near the quantum critical point
Eq.~(\ref{bulk15}) may be written in general forms as
\begin{mathletters}\label{bulk16}
\begin{equation}
h={\cal M}^{\delta} f_{h} (\delta\lambda {\cal M}^{-1/\beta},
(t/\lambda)^{1/\nu}{\cal M}^{-1/\beta}),
\end{equation}
or
\begin{equation}
{\cal M}=\left(\frac{t}{\lambda}\right)^{-\beta/\nu} f_{{\cal M}}
(\delta\lambda {\cal M}^{-1/\beta},h{\cal M}^{-\delta}).
\end{equation}
\end{mathletters}

In Eqs.~(\ref{bulk16}) $f_{h}(x,y)$ and $f_{{\cal M}}(x,y)$ are some
scaling functions, furthermore $\gamma=\frac{2}{d-1}$,
$\nu=\frac{1}{d-1}$, $\beta=\frac{1}{2}$ and $\delta=\frac{d+3}{d-1}$
are the familiar bulk critical exponents for the $(d+1)$--dimensional
classical spherical model. Eqs.~(\ref{bulk16}) are direct verification
of FSS hypothesis in conjunction with classical to quantum critical
dimensional crossover. They can be easily transformed
into the scaling form (Eq. (21)) obtained in
Ref.~\onlinecite{vojta95}, however here they are verified for $1<d<3$
instead of $2<d<3$ (c.f. Ref.~\onlinecite{vojta95}), i.e. the
non--critical case is included.

Hereafter we will try to give an explicit expression of the scaling
function $f_{h}(x,y)$ ($x\equiv\delta\lambda{\cal M}^{-2}$,
$y\equiv(t/\lambda)^{d-1}{\cal M}^{-2}$) in the neighborhood of the
quantum critical fixed point. This may be performed, in the case
$\frac{t}{\lambda}\sqrt{h/{\cal M}}\ll1$, with the use of the
asymptotic form of ${\cal K}(\nu,y)$
the following result for the scaling function
\begin{equation}
f_{h}(x,y)=\left[\frac{(4\pi)^{d/2}}{\Gamma\left(1-\frac{d}{2}
\right)} y^{-\nu}\left(1+x+\frac{1}{2\pi^{(d+1)/2}} \Gamma\left(
\frac{d-1}{2}\right)\zeta(d-1) y\right)\right]^{2/(d-2)}.
\end{equation}

For the special case $d=2$
the scaling function is given by the expression
\begin{equation}\label{bulk18}
f_{h}(x,y)= 4y^{2}\left[\text{arcsinh}\frac{1}{2}\exp\left(2\pi
\frac{1+x}{y}\right)\right]^{2}.
\end{equation}

At $x=0$ and $y\gg1$ (fixed low--temperature and
$h\rightarrow0^{+}$)
Eq.~(\ref{bulk18}) reduces to
\begin{equation}
f_{h}(0,y)\approx y^{2} \exp \left(\frac{4\pi}{y}\right).
\end{equation}
In the region $x<-1$, and for $y\ll1$ (fixed weak field and
$t\rightarrow0^{+}$) the corresponding scaling function is
\begin{equation}
f_{h}(x,y)\approx y^{2} \exp\left(4\pi\frac{x+1}{y}\right),
\end{equation}
and in region $x>-1$ and $y\ll1$ we have
\begin{equation}\label{bulk19}
f_{h}(x,y)\approx16\pi^{2}(x+1)^{2}\left[1+\frac{y}{\pi(1+x)}\exp\left(
-4\pi\frac{1+x}{y}\right)\right].
\end{equation}
This identifies the zero temperature ($y=0$) form of the
scaling function~(\ref{bulk19}) with those of the 3-dimensional
classical spherical model.

\section{System confined to a finite geometry}\label{fing}
When the model Hamiltonian~(\ref{model1}) is confined to the general
geometry $L^{d-d'}\!\!\times\!\!\infty^{d'}\!\!\times\!\!L_{\tau}$,
with $0\leq d' \leq d$,
equation~(\ref{model3}) of the spherical field $\phi$ takes the
form (for a derivational details see Appendix~\ref{app1})
\begin{eqnarray}
\frac{1}{\lambda}&=&\frac{1}{\lambda_{c}}- (4\pi)^{-(d+1)/2} \left|
\Gamma \left( \frac{1-d}{2} \right) \right|\phi^{(d-1)/2} \nonumber\\
& &+\frac{\phi^{(d-1)/2}}{(2\pi)^{(d+1)/2}} \sum_{m,
{\bbox l}(d-d')}^{\hskip7mm\prime} \frac{K_{(d-1)/2} \left[\phi^{1/2}
\left\{(\lambda m/t)^{2} + (L |{\bbox l}|)^{2}\right\}^{1/2}
\right]}{\left[\phi^{1/2} \left\{(\lambda m/t)^{2} +
(L|{\bbox l}|)^{2}\right\}^{1/2}\right]^{(d-1)/2}}
+\frac{h^{2}}{\phi^{2}}\label{finite},
\end{eqnarray}
where
$$
|{\bbox l}| = \left(l_{1}^{2}+l_{2}^{2}+\cdots+l_{d-d'}^{2}
\right)^{1/2}
$$
and the primed summation indicates that the vector with components
$m=l_{1}=l_{2}=\cdots=l_{d-d'}=0$ is excluded.

\subsection{Shift of the critical quantum parameter}\label{scqp}

The FSS theory (for a review see Ref.~\onlinecite{barber83}) asserts,
for the temperature driven phase transition, that the phase transition
occurring in the system at the thermodynamic limit persists, if the
dimension $d'$ of infinite sizes is greater than the lower critical
dimension of the system. In this case the value of the critical
temperature $T_{c}(\infty)$ at which some thermodynamic functions
exhibit a singularity is shifted to $T_{c} (L)$ critical temperature
for a system confined to the general geometry
$L^{d-d'}\!\!\times\!\!\infty^{d'}$, when the system is infinite in
$d'$ dimensions and finite in $(d-d')$--dimensions. In the case when
the number of infinite dimensions is less than the lower critical
dimension, there is no phase transition in the system and the
singularities of the thermodynamic functions are altered. The critical
temperature $T_{c}(\infty)$ in this case is shifted to a
pseudocritical temperature, corresponding to the center of the
rounding of the singularities of the thermodynamic functions, holding
in the thermodynamic limit.

In our quantum case, having in mind that we have considered the
low--temperature behavior of model~(\ref{model1}) in the context of
the FSS theory it is convenient to choose the quantum parameter
$\lambda$ like a critical instead of the temperature $t$ and to
consider our system confined to the geometry
$L^{d-d'}\!\!\times\!\!\infty^{d'}\!\!\times\!\!L_{\tau}$. So the
shifted critical quantum parameter
$\lambda_{c}(t,L)\equiv\lambda_{tL}$ is obtained by setting $\phi=0$
in Eq.~(\ref{finite}). This gives
\begin{equation}\label{shift}
\frac{1}{\lambda_{tL}}-\frac{1}{\lambda_{c}} =
\frac{\Gamma\left(\frac{d-1}{2}\right)}
{4\pi^{(d+1)/2}} \sum_{m,{\bbox l}(d-d')}^{\hskip7mm\prime}
\left((\lambda_{tL} m/t)^{2}+ (L |{\bbox l}|)^{2}\right)^{(1-d)/2}.
\end{equation}

The sum in the r.h.s. of Eq.~(\ref{shift}) is convergent for
$d'>2$, however it can be expressed in terms of the Epstein
zeta function 
\begin{equation}\label{epstein}
{\cal Z} \left| \begin{array}{c}
0\\0
\end{array} \right| \left(L^{2}{\bbox l}^{2}
+\left(\frac{\lambda}{t}\right)^{2}m^{2};d-1\right) =
\sum_{m,{\bbox l} (d-d')}^{\hskip7mm\prime}
\left(L^{2}{\bbox l}^{2}+\left(\frac{\lambda}{t}\right)^{2}m^{2}
\right)^{\frac{1-d}{2}},
\end{equation}
which can be regarded as the generalized ($d-d'+1$)-dimensional
analog of the Riemann zeta function $\zeta (\frac{d-1}{2})$ (see
Ref.~\onlinecite{glasser80}). In the case  under consideration the
Epstein zeta function has only a simple pole at $d' = 2$ and may be
analytically continued for $0 \leq d' < 2$ to give a meaning to
Eq.~(\ref{shift}) for $d'<2$ as well. It is hard to investigate the
sum appearing in Eq.~(\ref{epstein}). The anisotropy of the sum
$L^{2}l_{1}^{2}+\cdots+L^{2}l_{d-d'}^{2}+\left(\frac{\lambda}{t}
\right)^{2}m^{2}$ is an additional problem. That is for what we
will try to solve it asymptotically, considering different regimes
of the temperature, depending on that wether
$L\ll\frac{\lambda_{tL}}{t}$ or $L\gg\frac{\lambda_{tL}}{t}$ which
will be called, respectively, very low--temperature regime and
low--temperature regime.

\subsubsection{Low--temperature regime $\frac{\lambda_{tL}}{t}\ll L$}
In this case after some algebra the resulting expression is
\begin{eqnarray}
\frac{1}{\lambda_{tL}}-\frac{1}{\lambda_{c}} &=&\frac{1}{2
\pi^{(d+1)/2}}\left(\frac{t}{\lambda_{tL}}\right)^{d-1}
\Gamma\left(\frac{d-1}{2}\right)\zeta(d-1)
+\frac{t}{\lambda_{tL}}\frac{L^{2-d}}{4\pi^{d/2}}
\Gamma\left(\frac{d}{2}-1\right) \sum_{{\bbox l}(d-d')}^{\hskip7mm\prime}
|{\bbox l}|^{2-d} \nonumber \\
& &+\left(\frac{t}{\lambda_{tL}}\right)^{d/2}\frac{L^{1-d/2}}{\pi}
\sum_{{\bbox l}(d-d')}^{\hskip7mm\prime}\sum_{m=1}^{\infty}
\left(\frac{m}{|{\bbox l}|}\right)^{d/2-1}K_{\frac{d}{2}-1}
\left(2\pi\frac{t}{\lambda_{tL}}Lm|{\bbox l}| \right).
\label{shift1}
\end{eqnarray}

The first term of the rhs of Eq.~(\ref{shift1}) is the shift of the
critical quantum parameter (see Eq.~(\ref{bulk10})) due
to the presence of the quantum effects in
the system. The second term is a correction resulting from the finite
sizes. It is just the shift due the finite--size effects in the
$d$--dimensional spherical model~\cite{chamati96} multiplied by the
temperature scaled to the quantum parameter. Here the $(d-d')$--fold
sum may be continued analytically beyond its domain of convergence
with respect to "$d$" and "$d'$" (which is $2<d'<d$). The last term is
exponentially small in the considered limit i.e.
$\frac{\lambda_{tL}}{t}\ll L$.

In the borderline case $d=2$, Eq.~(\ref{shift1}) reduces to
\begin{equation}\label{gammae}
\frac{1}{\lambda_{tL}}-\frac{1}{\lambda_{c}}=\frac{t}{2\pi\lambda_{tL}}
\left\{B_{0}+\gamma_{E}^{}+\ln\frac{tL}{2\lambda_{tL}}+2 \sum_{{\bbox l}
(2-d')}^{\hskip7mm\prime}\sum_{m=1}^{\infty}
K_{0}\left(2\pi\frac{t}{\lambda_{tL}}Lm|{\bbox l}| \right)\right\},
\end{equation}
where $\gamma_{E}^{}=.577...$ is the Euler constant and $B_{0}$ is a
constant depending on the dimensionality $d-d'$.

In the particular case of strip geometry ($d'=1$), $B_{0}=-\ln2\pi$ and
the shift is given by
\begin{equation}\label{lt<1}
\frac{1}{\lambda_{tL}}-\frac{1}{\lambda_{c}}\approx\frac{t}{2\pi\lambda_{tL}}
\left\{\gamma_{E}^{}+\ln\frac{tL}{4\pi\lambda_{tL}}\right\}.
\end{equation}

For the fully finite geometry case ($d'=0$), $B_{0}=-\ln\frac{\left[\Gamma
\left(\frac{1}{4}\right)\right]^{2}}{2\sqrt{\pi}}$ and
we obtain for the shift
\begin{equation}\label{tl<1}
\frac{1}{\lambda_{tL}}-\frac{1}{\lambda_{c}}\approx\frac{t}{2\pi
\lambda_{tL}}\left\{\gamma_{E}^{}+\ln\frac{tL}{4\pi\lambda_{tL}}-
\ln\frac{\left[\Gamma\left(\frac{1}{4}\right)\right]^{2}}{4\pi
\sqrt{\pi}}\right\}.
\end{equation}
Let us note that in the rhs of Eqs.~(\ref{lt<1}) and~(\ref{tl<1})
exponentially small corrections are omitted.
            
\subsubsection{Very low--temperature regime $\frac{\lambda}{t}\gg L$}
The final result in this case is given by the expression
\begin{eqnarray}
\frac{1}{\lambda_{tL}}-\frac{1}{\lambda_{c}} &=&
\frac{L^{1-d}}{4\pi^{(d+1)/2}} \Gamma\left(\frac{d-1}{2}\right)
\sum_{{\bbox l}(d-d')}^{\hskip7mm\prime} |{\bbox l}|^{1-d}
+\frac{L^{d'-d}}{2\pi^{(d'+1)/2}}\Gamma\left(\frac{d'-1}{2}\right)
\left(\frac{t}{\lambda_{tL}}\right)^{d'-1}\zeta(d'-1)\nonumber\\
& &+\frac{L^{1/2+d'/2-d}}{\pi} \left(\frac{t}{\lambda_{tL}}
\right)^{\frac{d'-1}{2}}\sum_{{\bbox l}(d-d')}^{\hskip7mm\prime}
\sum_{m=1}^{\infty}\left(\frac{|{\bbox l}|}{m}\right)^{
\frac{d'-1}{2}}K_{\frac{d'-1}{2}}
\left(2\pi\frac{\lambda_{tL}}{tL}m|{\bbox l}| \right).
\label{shift2}
\end{eqnarray}
Here, in the rhs, the first term is the expression of the shift of the
critical quantum parameter, at  zero temperature,~\cite{chamati94} due
to the finite sizes of the system. This is equivalent to the shift of
a ($d+1$)--dimensional spherical model confined to the geometry
$L^{d+1-d'}\!\!\times\!\!\infty^{d'}$. The second term gives a
corrections due to the quantum effects. This is the shift of critical
quantum parameter of a $d'$--dimensional infinite system multiplied by
the volume of a $(d-d')$--dimensional hypercube. The third term is
exponentially small in the limit of very low temperatures.
For $d'=1$ Eq.~(\ref{shift2}) yields
\begin{equation}\label{d'=1}
\frac{1}{\lambda_{tL}}-\frac{1}{\lambda_{c}}=\frac{L^{1-d}}{2\pi}
\left\{\ln\frac{\lambda_{tL}}{2\sqrt{\pi}tL}+\frac{\gamma_{E}^{}}{2}+
\tilde{C_{0}}+2\sum_{{\bbox l}(d-1)}^{\hskip7mm\prime}
\sum_{m=1}^{\infty}K_{0}\left(2\pi\frac{\lambda_{tL}}{tL}m|\
{\bbox l}| \right)
\right\},
\end{equation}
where
$$
\tilde{C_{0}}=\left[2\pi^{(d-1)/2}\right]^{-1}
\Gamma\left(\frac{d-1}{2}\right) C_{0},
$$
and the expressions for the constants $C_{0}$ are quite complicated
expect for some special cases (see Refs.~\onlinecite{singh89}, e.g.
for $d=2$, $d'=1-\varepsilon$ we have $C_{0}=\gamma_{E}^{}-\ln(4\pi)$
(c.f. Ref.~\onlinecite{Zin89}, Eq. (30.104)).

In the particular case $d'=1$ and $d=2$ the constant
$C_{0}$ is given by $C_{0}=\gamma_{E}^{}-\ln4\pi$ and
\begin{equation}\label{tl>1}
\frac{1}{\lambda_{tL}}-\frac{1}{\lambda_{c}}\approx\frac{1}{2\pi L}\left
\{\gamma_{E}^{}+\ln\frac{\lambda_{tL}}{4\pi tL}\right\}.
\end{equation}

Comparing between Eqs.~(\ref{lt<1}) and~(\ref{tl>1}) one can see the
crucial role (in symmetric form) of $L$ or $\lambda_{tL}/t$ in the
low--temperature regime and the very low--temperature one,
respectively.

In the other particular case of a two--dimensional bloc geometry
$d'=0$ and $d=2$, from Eq.~(\ref{shift2}) one gets (up to
exponentially small corrections)
\begin{equation}\label{dp=0}
\frac{1}{\lambda_{tL}}-\frac{1}{\lambda_{c}}\approx\frac{L^{-1}}{\pi}
\zeta\left(\frac{1}{2}\right)\beta\left(\frac{1}{2}\right)
-\frac{\lambda_{tL}}{tL^{2}}\zeta(-1),
\end{equation}
where
$$
\beta(s)=\sum_{l=1}^{\infty}\frac{(-1)^{l}}{(2l+1)^{s}}.
$$

Instead of the previous case of low--temperature regime here the lower
quantum critical dimension $d'=1$ is responsible for the logarithmic
dependence in Eq.~(\ref{d'=1}). This is the reason for the significant
difference between Eqs.~(\ref{tl>1}) and~(\ref{dp=0}).

The obtained equations for $\lambda_{tL}$ will be exploited
later for the study of the two dimensional case.

\subsection{Zero--field susceptibility}\label{zerofs}
It is possible to transform Eq.~(\ref{finite}) in the following
equivalent forms
\begin{mathletters}\label{fssl}
\begin{equation}
L^{d-1}\delta\lambda+\left(\frac{h L^{(d+3)/2}}{L^{2} \phi}
\right)^{2}=\frac{(L\phi^{1/2})^{d-1}}{(4\pi)^{(d+1)/2}}
\left[\left|\Gamma\left(\frac{1-d}{2}\right)\right|-
2{\cal K}_{\frac{\lambda}{tL}}\left.\left(\frac{d-1}{2}\right|
d-d'+1,\frac{L\phi^{1/2}}{2}\right)\right],
\end{equation}
where
\begin{equation}\label{kat}
{\cal K}_{a} (\nu|p,y) = \sum_{m,{\bbox l}(p-1)}^{\hskip7mm\prime}
\frac{K_{\nu}\left(2y\sqrt{{\bbox l}^{2}+a^{2}m^{2}}\right)}
{\left(y\sqrt{{\bbox l}^{2}+a^{2}m^{2}}\right)^{\nu}}, \ \ y>0, \ \
{\bbox l}^2 = l_{1}^{2}+l_{2}^{2}+\cdots+l_{p-1}^{2}.
\end{equation}
\end{mathletters}
or
\begin{mathletters}\label{fsst}
\begin{eqnarray}
\left(\frac{t}{\lambda}\right)^{1-d}\delta\lambda
+\left[\frac{h}{\phi}\left(\frac{t}{\lambda}\right)^{2}
\left(\frac{\lambda}{t}\right)^{(d+3)/2}\right]^{2}&=&
\frac{(\lambda\phi^{1/2}/t)^{d-1}}{(4\pi)^{(d+1)/2}}
\left[\left|\Gamma\left(\frac{1-d}{2}\right)\right|\right.\nonumber\\
& &\left.-2\tilde{\cal K}_{\frac{tL}{\lambda}}
\left.\left(\frac{d-1}{2}\right|d-d'+1,\frac{\lambda\phi^{1/2}}{2t}
\right)\right],
\end{eqnarray}
where
\begin{equation}\label{katt1}
\tilde{\cal K}_{a} (\nu|p,y) = {\cal K}_{\frac{1}{a}} (\nu|p;ay)
=\sum_{m,{\bbox l}(p-1)}^{\hskip7mm\prime}
\frac{K_{\nu}\left(2y\sqrt{a^{2}{\bbox l}^{2}+m^{2}}\right)}
{\left(y\sqrt{a^{2}{\bbox l}^{2}+m^{2}}\right)^{\nu}}, \ \ y>0.
\end{equation}
\end{mathletters}
The functions ${\cal K}_{a} (\nu|p;y)$ and $\tilde{\cal K}_{a}
(\nu|p;y)$ are anisotropic generalizations of the ${\cal K}$--function
introduced in Ref.~\onlinecite{singh85}.

Eqs.~(\ref{fssl}) and~(\ref{fsst}) show that the correlation length
$\xi=\phi^{-1/2}$ will scale like
\begin{mathletters}
\begin{equation}
\xi=L f_{\xi}^{L} \left\{\delta\lambda L^{1/\nu},\frac{tL}{\lambda},
hL^{\Delta/\nu}\right\},
\end{equation}
or like
\begin{equation}
\xi=\frac{\lambda}{t}f_{\xi}^{t} \left\{\delta\lambda\left(\frac{t}
{\lambda}\right)^{-1/\nu},\frac{tL}{\lambda},h\left(\frac{t}{\lambda}
\right)^{-\Delta/\nu}\right\},
\end{equation}
\end{mathletters}
which suggests also that there will be some kind of interplay
(competition) between the finite--size and the quantum effects.

Hereafter we will try to find the behavior of the susceptibility
$\chi=\phi^{-1}$
as a function of the temperature $t$ and the size $L$ of the system.
For simplicity, in the remainder of this section, we will
investigate the free field case ($h=0$).

{\bf 1)} For $\frac{\lambda}{t}\phi^{1/2}\ll1$, after using the
asymptotic form of the function defined in~(\ref{kat})
(see Appendix~\ref{app2}) Eq.~(\ref{finite}) reads ($d'\neq2$,
$1<d<3$)
\begin{equation}\label{fssa}
\delta\lambda+\frac{t}{\lambda}\frac{L^{d'-d}}{(4\pi)^{d'/2}}
\Gamma\left(1-\frac{d'}{2}\right)\phi^{(d'-2)/2}+\frac{1}
{4\pi^{(d+1)/2}}\Gamma\left(\frac{d-1}{2}\right)
\sum_{m,{\bbox l}(d-d')}^{\hskip7mm\prime}
\left[\left(\frac{\lambda}{t}m\right)^{2}+\left(L{\bbox l}\right)^{2}
\right]^{\frac{1-d}{2}}=0.
\end{equation}

Now we will examine Eq.~(\ref{fssa}) in different regimes of $t$ and
$L$ and for different geometries of the lattice:

{\bbox a}. $\frac{\lambda}{t}\phi^{1/2}\ll1$ and
$\frac{tL}{\lambda}\gg1$: In this case Eq.~(\ref{fssa}) transforms
into (up to a exponentially small corrections c.f. with
Eq.~(\ref{shift1}))
\begin{eqnarray}
0&=&\delta\lambda+\frac{t}{\lambda}\frac{L^{d'-d}}{(4\pi)^{d'/2}}
\Gamma\left(1-\frac{d'}{2}\right)\phi^{(d'-2)/2}+
\frac{1}{2\pi^{(d+1)/2}}\left(\frac{t}{\lambda}\right)^{d-1}
\Gamma\left(\frac{d-1}{2}\right)\zeta(d-1)\nonumber\\
& &+\frac{t}{\lambda}\frac{L^{2-d}}{4\pi^{d/2}}\Gamma\left(\frac{d}{2}
-1\right)\sum_{{\bbox l}(d-d')}^{\hskip7mm\prime} |{\bbox l}|^{2-d}.
\label{lta}
\end{eqnarray}
This equation has different type of solutions depending on that
whether the dimensionality $d$ is above or below the classical
critical dimension 2.

At $\lambda=\lambda_{c}$ and when $d'<2<d<3$ (i.e. when there is no
phase transition in the system) we obtain for the zero--field
susceptibility
\begin{equation}\label{ltra}
\chi=\left(\frac{t}{\lambda_{c}}\right)^{-2}
\left(\frac{tL}{\lambda_{c}}\right)^{2(d-d')/(2-d')}
\left[\frac{2^{d'-1}}{\pi^{(d-d'+1)/2}}\frac{\Gamma\left(
\frac{d-1}{2}\right)}{\Gamma\left(1-\frac{d'}{2}\right)}
\zeta(d-1)\right]^{\frac{2}{2-d'}}.
\end{equation}
However for $1<d<2$, Eq.~(\ref{lta}) has no solution at
$\lambda=\lambda_{c}$ obeying the initial condition
$\frac{\lambda_{c}}{t}\phi^{1/2}\ll1$.

Eq.~(\ref{ltra}) generalizes the bulk result~(\ref{bulk8}) for $d$
close to the upper quantum critical dimension i.e. $d=3$.

At the shifted critical quantum parameter $\lambda_{c}(t)$ given by
Eq.~(\ref{bulk10}) we get
\begin{equation}\label{zin}
\chi=L^{2} \left[\frac{2^{d'-2}}{\pi^{(d-d')/2}}
\frac{\Gamma\left(\frac{d}{2}-1\right)}{\Gamma\left(1-\frac{d'}{2}
\right)}\sum_{{\bbox l}(d-d')}^{\hskip7mm\prime} |{\bbox l}|^{2-d}
\right]^{\frac{2}{2-d'}}.
\end{equation}
However this solution is valid only for $3>d>2>d'$ i.e. here again
there is no phase transition in the system.

{\bbox b}. $\frac{\lambda}{t}\phi^{1/2}\ll1$ and
$\frac{tL}{\lambda}\ll1$: In this case, Eq.~(\ref{fssa})
gives (up to exponentially small corrections c.f. with
Eq.~(\ref{shift2}))
\begin{eqnarray}
0&=&\delta\lambda+\frac{t}{\lambda}\frac{L^{d'-d}}{(4\pi)^{d'/2}}
\Gamma\left(1-\frac{d'}{2}\right)\phi^{(d'-2)/2}+
\frac{L^{1-d}}{4\pi^{(d+1)/2}} \Gamma\left(\frac{d-1}{2}\right)
\sum_{{\bbox l}(d-d')}^{\hskip7mm\prime}|{\bbox l}|^{1-d}\nonumber \\
& &+\frac{L^{d'-d}}{2\pi^{(d'+1)/2}}\Gamma\left(\frac{d'-1}{2}\right)
\left(\frac{t}{\lambda}\right)^{d'-1}\zeta(d'-1).\label{vlta}
\end{eqnarray}
Here we find that the solutions of Eq.~(\ref{vlta}) depend upon
that whether the dimensionality $d'<1$ or $d'>1$.

At $\lambda=\lambda_{c}$ and for $1<d'<2$, Eq.~(\ref{vlta}) has
\begin{equation}\label{vltra}
\chi=L^{2}\left(\frac{\lambda_{c}}{tL}\right)^{2/(2-d')}
\left[\frac{2^{d'-2}}{\pi^{(d-d'+1)/2}}
\frac{\Gamma\left(\frac{d-1}{2}\right)}{\Gamma\left(1-\frac{d'}{2}
\right)}\sum_{{\bbox l}(d-d')}^{\hskip7mm\prime} |{\bbox l}|^{1-d}
\right]^{\frac{2}{2-d'}}
\end{equation}
as a solution. For $0\leq d'<1$, however, it has no solution obeying
the initially imposed restriction $\frac{\lambda_{c}}{t}\phi^{1/2}\ll1$.

At the shifted critical quantum parameter $\lambda_{c}(L)$ given
by~\cite{chamati94}
$$
\frac{1}{\lambda}-\frac{1}{\lambda_{c}(L)}=\frac{L^{1-d}}
{4\pi^{(d+1)/2}}\Gamma\left(\frac{d-1}{2}\right)
\sum_{{\bbox l}(d-d')}^{\hskip7mm\prime} |{\bbox l}|^{1-d},
$$
Eq.~(\ref{vlta}) has a solution obeying the initial condition
$\frac{\lambda_{c}}{t}\phi^{1/2}\ll1$ only for $d'=1+\varepsilon$ and
in this case the susceptibility behaves like
\begin{equation}\label{epst}
\chi=\frac{1}{(\pi\varepsilon)^{2}}\frac{\lambda_{c}^{2}}{t^{2}}
\left[1-\varepsilon\left(\gamma_{E}^{}+\ln\frac{\varepsilon}
{2}\right)\right]^{2}.
\end{equation}

{\bf 2)} For $L\phi^{1/2}\ll1$, from Eqs.~(\ref{fssl}) and
Eq.~(\ref{lastap}) we get once again Eq.~(\ref{lta}).
In spite of the fact that we have the same equation as in the case
$\frac{\lambda}{t}\phi^{1/2}\ll1$, the expected solutions for the
susceptibility may be different because of the new imposed condition.
Here also we will consider the two limiting cases of low--temperature
and very low--temperature regimes.

{\bbox a}. $L\phi^{1/2}\ll1$ and $\frac{tL}{\lambda}\gg1$: In this case
Eq.~(\ref{fssa}) again is transformed into Eq.~(\ref{lta}) and we
obtain at $\lambda=\lambda_{c}$ the solution given by
Eq.~(\ref{ltra}), which is valid only for $d'<2<d<3$, i.e. we have the
same solution as in the previous case i.e.
$\frac{\lambda}{t}\phi^{1/2}\ll1$.

At $\lambda=\lambda_{c}(t)$, we formally obtain Eq.~(\ref{zin})
which, however, may be considered like a solution  only in the
neighborhood of the lower classical critical dimension $d=2$. For the
cylindric geometry ($d'=1$ and $d=2+\varepsilon$) we get
\begin{equation}\label{cy}
\chi=\frac{L^{2}}{(\pi\varepsilon)^{2}}\left[1-\frac{\varepsilon}{2}
(\gamma_{E}^{}-\ln4\pi)\right]^{2}.
\end{equation}
This result is contained in Eq. (30.109) of Ref.~\onlinecite{Zin89} in
the large $n$--limit case for the NLQ$\sigma$M.

In the case of slab geometry $d-d'=1$ ($d=2+\varepsilon,
d'=1+\varepsilon$) instead of~(\ref{cy}) we obtain
\begin{equation}\label{cy1}
\chi=\frac{L^{2}}{(\pi\varepsilon)^{2}}\left[1-\varepsilon(
\gamma_{E}^{}-\ln2)-\varepsilon\ln\varepsilon\right]^{2}.
\end{equation}

In the case of a bloc geometry ($d=2+\varepsilon$ and $d'=0$) we find
the following behavior for the susceptibility
\begin{equation}
\chi=\frac{L^{2}}{2\pi\varepsilon}\left[1-\frac{\varepsilon}{4}\left(
\gamma_{E}^{}-\ln\frac{\left[\Gamma\left(\frac{1}{4}\right)\right]^{4}}
{4\pi^{2}}\right)\right]^{2}.
\end{equation}

For the case of "quasi--bloc geometry" ($d=2+\varepsilon$ and
$d'=\varepsilon$) we get
\begin{equation}\label{epsb}
\chi=\frac{L^{2}}{2\pi\varepsilon}\left[1-\frac{\varepsilon}{4}\left(
2\gamma_{E}^{}+\ln\frac{\pi\varepsilon}{2}-2\ln\frac{\left[\Gamma\left(
\frac{1}{4}\right)\right]^{2}}{2\sqrt{\pi}}\right)\right]^{2}.
\end{equation}
Here the appearance of $\varepsilon$ in the denominator in
formulas~(\ref{epst})--(\ref{epsb}) signalizes that the scaling in its
simple form will fail at $\varepsilon=0$.

{\bbox b}. $L\phi^{1/2}\ll1$ and $\frac{tL}{\lambda}\ll1$: Here we find
that Eq.~(\ref{vlta}) is valid, and it has Eq.~(\ref{vltra}) as a
solution at $\lambda=\lambda_{c}$ and for $0\leq d'<1$. For $1<d'<2$
the susceptibility is given by
\begin{equation}\label{vltrb}
\chi=\left(\frac{\lambda_{c}}{2t}\right)^{2} \left[\frac{2}{\pi^{1/2}}
\frac{\Gamma\left(\frac{d'-1}{2}\right)}{\Gamma\left(1-\frac{d'}{2}
\right)}\zeta(d'-1)\right]^{\frac{2}{2-d'}}.
\end{equation}

At the shifted critical point $\lambda_{c}(L)$, for the susceptibility
we obtain Eq.~(\ref{vltrb}) under the restriction $2>d'>1$, which
guarantees the positiveness of the quantity under brackets.

When $\lambda<\lambda_{c}$ for $1<d<3$ and $d'<2$, i.e. when
there is no phase transition in the system, we obtain
\begin{equation}\label{bussa}
\chi=\left[\frac{(4\pi)^{d'/2}}{\Gamma\left(1-\frac{d'}{2}\right)}
\left(1-\frac{\lambda}{\lambda_{c}}\right)\right]
^{\frac{2}{2-d'}} t^{-2/(2-d')} L^{2(d-d')/(2-d')}.
\end{equation}

If $d'>2$ there is a phase transition in the system at the shifted
value of the critical quantum parameter $\lambda_{tL}$ (the shift in
this case is due to the quantum and finite--size effects) and
Eq.~(\ref{fssa}) transforms to
\begin{equation}
1-\frac{\lambda}{\lambda_{tL}}=t\frac{L^{d'-d}}{(4\pi)^{d/2}}
\Gamma\left(1-\frac{d'}{2}\right)\phi^{(d'-2)/2},
\end{equation}
which has the following solutions
\begin{equation}\label{bussb}
\chi=\left\{
\begin{array}{ll}
\left[\frac{(4\pi)^{d'/2}}{\Gamma\left(1-\frac{d'}{2}\right)}
\left(1-\frac{\lambda}{\lambda_{tL}}\right)\right]
^{\frac{2}{2-d'}} t^{-2/(2-d')} L^{2(d-d')/(2-d')},
& \lambda>\lambda_{tL}\\
\infty ,
& \lambda\leq\lambda_{tL}
\end{array}
\right.
\end{equation}
Let us notice that Eqs.~(\ref{bussa}) and~(\ref{bussb}) are the
finite--size forms, for the susceptibility, of Eqs.~(\ref{suscepb})
and~(\ref{bulk11}), respectively, found for the bulk system.

\subsection{Two--dimensional case}\label{tdc}
The two--dimensional case needs special treatment because of its
physical reasonability and the increasing interest in the context of
the quantum critical phenomena.~\cite{chakravarty89,%
chubukov94,sachdev94,Rosenstein90,%
Castro93,fujii95} From Eq.~(\ref{finite}) for $d=2$ and in
the absence of a magnetic field $h=0$ we get
\begin{equation}\label{2dcase}
\delta\lambda=\frac{\phi^{1/2}}{4\pi}-\frac{1}{4\pi}
\sum_{m,{\bbox l}(2-d')}^{\hskip7mm\prime}
\frac{\exp\left[-\phi^{1/2}\left(\frac{\lambda^{2}}{t^{2}}m^{2} +
L^{2}{\bbox l}^{2}\right)^{1/2}\right]}{\left(\frac{\lambda^{2}}
{t^{2}}m^{2} + L^{2}{\bbox l}^{2}\right)^{1/2}}.
\end{equation}
Introducing the scaling functions
$Y_{t}^{d'}=\frac{\lambda}{t}\phi^{1/2}$ and $Y_{L}^{d'}=L\phi^{1/2}$,
where the superscript $d'$ denotes the number of infinite
dimensions in the system, and the scaling variable
$a=\frac{tL}{\lambda}$ it is easy to write Eq.~(\ref{2dcase}) in the
scaling forms given in Eqs.~(\ref{fssl}) and (\ref{fsst}). The
solutions of the obtained scaling equations will depend on the number
of the infinite dimensions in the system. Here we will consider the
two most important particular cases:  strip geometry $d'=1$ and bloc
geometry $d'=0$. Our analysis will be confined to the study of the
behavior of the scaling functions at the critical value of the
quantum parameter $\lambda_{c}$, and at the shifted critical quantum
parameter $\lambda_{tL}$ (see Section~\ref{scqp}). It is difficult to
solve Eq~(\ref{2dcase}) by using an analytic approach that is for what
we will give a numerical treatment of the problem. It is, however,
possible to consider the two limits: $a\gg1$ i.e. the
low--temperature regime and $a\ll1$ i.e. the very low--temperature
regime.

\begin{figure}
\epsfxsize=4.5in
\centerline{\epsffile{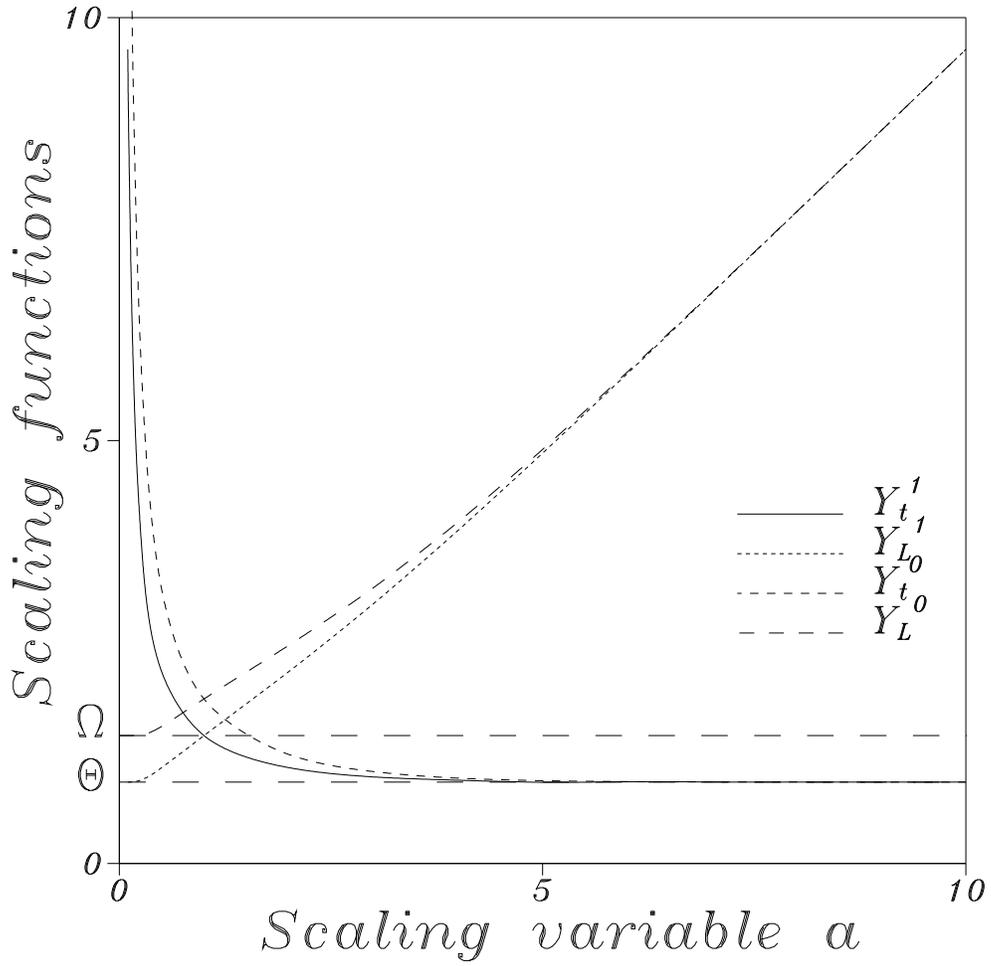}}
\vspace{0.1in}
\caption{The effects of the finite--size geometry on the bulk behavior
of $\phi^{1/2}$ for the two dimensional case at $\lambda=\lambda_{c}$.
The superscript $d'$ in $Y^{d'}_{L}=L\phi^{1/2}$ and
$Y^{d'}_{t}=\frac{\lambda_{c}}{t}\phi^{1/2}$ indicates the number of
infinite dimensions in the system. The scaling variable
$a=\frac{tL}{\lambda_{c}}$. The universal numbers are
$\Theta=0.962424...$ (see Eq.~(\protect\ref{theta})) and
$\Omega=1.511955...$}\label{fig2}
\end{figure}

{\it Strip geometry ($d'=1$)}:
In this case in the rhs of Eq.~(\ref{2dcase}) we have a two--fold sum
which permits a numerical analysis of the geometry under
consideration. FIG.~\ref{fig2} graphs the variation of the scaling
functions $Y_{t}^{1}$ and $Y_{L}^{1}$ against the variable $a$ at
$\lambda=\lambda_{c}$. This shows that for comparatively small value
of the scaling variable $a\sim5$ the finite--size behavior (see the
curve of the function $Y_{t}^{1}(a)$) merges in the low temperature
bulk one, while the behavior of $Y_{L}^{1}(a)$ shows that for
relatively not very low--temperatures ($a\sim\frac{1}{5}$, $L$-fixed)
the system simulates the behavior of a three--dimensional classical
spherical model with one finite dimensions. The mathematical reasons
for this are the exponentially small values of the corrections, as we
will show bellow.

{\it Bloc geometry ($d'=0$)}: In this case the three--fold sum in the
rhs of Eq.~(\ref{2dcase}) is not an obstacle to analyze it
numerically. For $\lambda=\lambda_{c}$ the behavior of the scaling
functions $Y_{L}^{0}(a)$ and $Y_{t}^{0}(a)$ is presented in
FIG.~\ref{fig2}. They have the same qualitative behavior as in
the strip geometry the only difference is the appearance of a new
universal number for $t=0$, i.e. $\Omega$, instead of the constant
$\Theta$ as a consequence of the asymmetry of the sum in the
low--temperature and the very low--temperature regimes.

Analytically for arbitrary values of the number of infinite dimensions
$d'$ we can treat first the problem  in the low--temperature regime
($a\gg1$). In this limit Eq.~(\ref{2dcase}) can be transformed into
(up to small corrections ${\cal O}(e^{-2\pi a})$)
\begin{equation}\label{2da>1}
\delta\lambda=\frac{1}{2\pi\lambda}\ln2\sinh
\frac{\lambda}{2t}\phi^{1/2}-\frac{1}{2\pi\lambda}\sum_{{\bbox l}(2-d')}
^{\hskip7mm\prime}K_{0}\left(L\phi^{1/2}|{\bbox l}|\right).
\end{equation}
For $\lambda=\lambda_{c}$ Eq.~(\ref{2da>1}) has the solution
\begin{equation}
\chi^{-1/2}\approx\frac{t}{\lambda_{c}}\Theta +
(2-d')\sqrt{\frac{2\pi}{5\Theta}}\left(\frac{t}{L\lambda_{c}}
\right)^{1/2}\exp\left(-\frac{tL}{\lambda_{c}}\Theta\right)
\end{equation}
i.e. the finite--size corrections to the bulk behavior are
exponentially small.

In the very low--temperature regime ($a\ll1$), Eq.~(\ref{2dcase})
reads (up to ${\cal O}(e^{-2\pi/a})$)
\begin{equation}\label{2da<1}
\delta\lambda=\frac{\phi^{1/2}}{4\pi}-\frac{L^{-1}}{4\pi}\sum_{{\bbox l}
(2-d')}^{\hskip7mm\prime}
\frac{\exp\left(-L\phi^{1/2}|{\bbox l}|\right)}{|{\bbox l}|}+
\frac{L^{d'-2}}{\pi^{(d'+1)/2}}\left(2\frac{\lambda}{t}\right)
^{\frac{1-d'}{2}}\sum_{m=1}^{\infty}K_{\frac{d'-1}{2}}
\left(\frac{\lambda}{t}\phi^{1/2}\right),
\end{equation}
which has the solutions:
\begin{equation}
\chi^{-1/2}\approx\frac{1}{L}\Theta +
\sqrt{\frac{2\pi}{5\Theta}}\left(\frac{L\lambda_{c}}{t}\right)^{1/2}
\exp\left(-\frac{\lambda_{c}}{tL}\Theta\right)
\end{equation}
for $d'=1$, and
\begin{equation}\label{Omega}
\chi^{-1/2}\approx\frac{1}{L}\Omega+\frac{1}{L}\left\{\frac{1}{2\Omega}
+\frac{\Omega}{2}\sum_{{\bbox l}(2)}^{\hskip7mm\prime} \left(\Omega^{2}+
4\pi^{2}{\bbox l}^{2}\right)^{-3/2}\right\}^{-1}\exp\left(-\Omega\frac
{\lambda_{c}}{tL}\right)
\end{equation}
for $d'=0$. Here $\Omega=1.511955...$ is an universal constant.

\begin{figure}
\epsfxsize=4.5in
\centerline{\epsffile{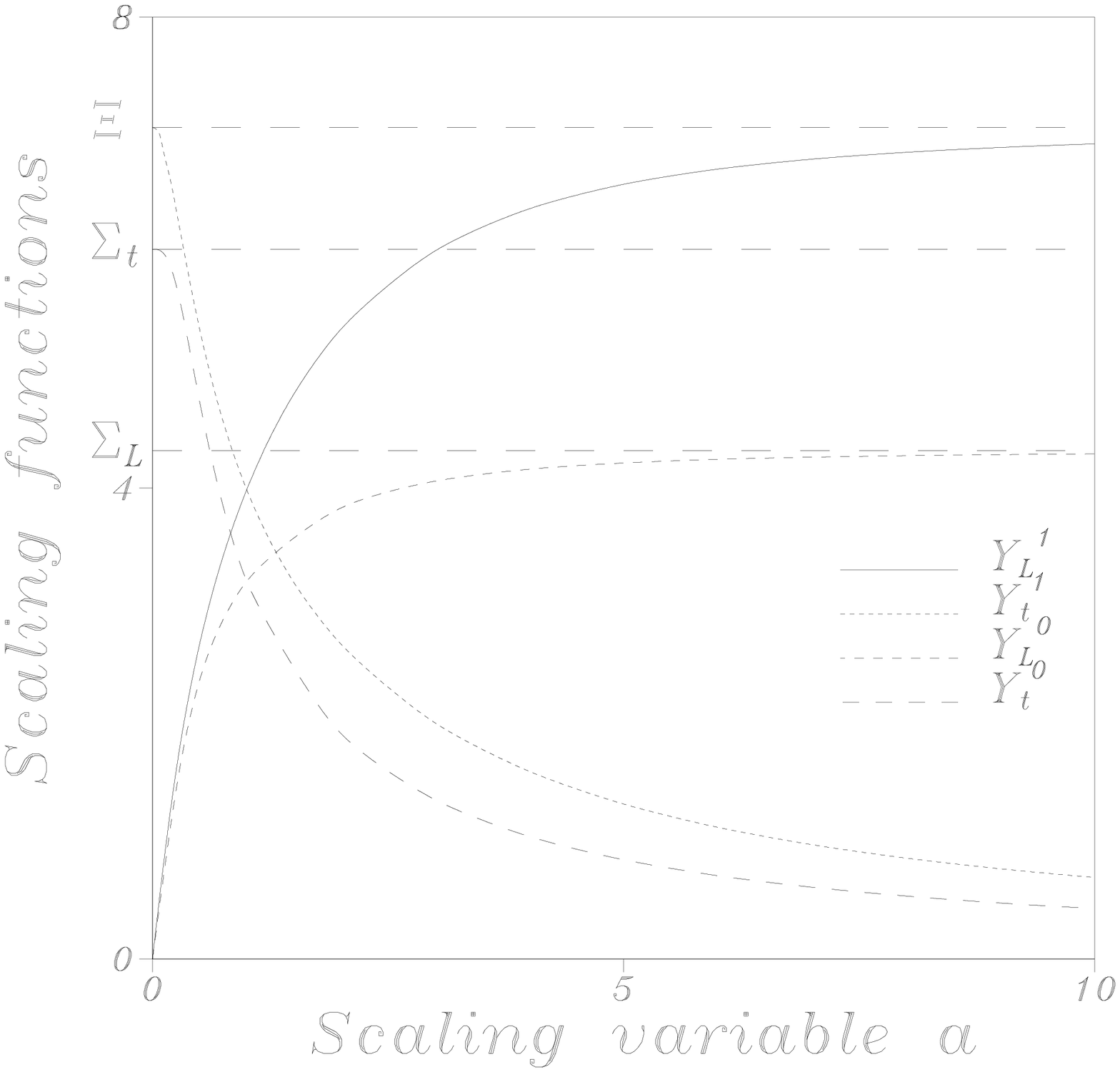}}
\vspace{0.1in}
\caption{The same as in FIG.~\protect\ref{fig2} but for
$\lambda=\lambda_{tL}$ and $a=\frac{tL}{\lambda_{tL}}$. The universal
numbers are: $\Xi=7.061132...$, $\Sigma_{t}=6.028966...$ and
$\Sigma_{L}=4.317795...$.}\label{fig3}
\end{figure}

In Section~\ref{scqp} an analytic continuation of the shift of the
critical quantum parameter for $d=2$ was presented. It is possible to
consider the solutions of Eq.~(\ref{2dcase}) at $\lambda=\lambda_{tL}$
(from Eqs.~(\ref{lt<1}),~(\ref{tl<1}),~(\ref{tl>1}) and~(\ref{dp=0}))
and for different geometries. In this case the scaling functions
$Y_{t}^{1}$, $Y_{L}^{1}$, $Y_{t}^{0}$ and $Y_{L}^{0}$ are graphed in
FIG.~\ref{fig3}. For $d'=1$ again we see that a symmetry between the
two limits $a\ll1$ and $a\gg1$ take place, since the scaling functions
$Y_{t}^{1}$ and $Y_{L}^{1}$ are limited by the universal constant
$\Xi$. The asymmetric case $d'=0$ has two different constants
$\Sigma_{t}$ and $\Sigma_{L}$, limiting the solutions of $Y_{t}^{0}$
and $Y_{L}^{1}$ from above.

The constants $\Xi$, $\Sigma_{t}$ and $\Sigma_{L}$ are obtained from
the asymptotic analysis (with respect to $a$) of Eq.~(\ref{2dcase})
for $\lambda=\lambda_{tL}$.

In the limit $a\gg1$ for arbitrary values of $d'$ we get (from
Eq.~(\ref{2da>1}))
\begin{equation}\label{a>>1}
B_{0}+\gamma_{E}^{}+\ln\frac{L\phi^{1/2}}{2}=
\sum_{{\bbox l}(2-d')}^{\hskip7mm\prime}
K_{0}\left(L\phi^{1/2}|{\bbox l} |\right),
\end{equation}
where the equation of $\lambda_{tL}$ from Eq.~(\ref{gammae}) is used.
Eq.~(\ref{a>>1}) has the solutions:
\begin{equation}\label{XiSigma}
L\chi^{-1/2}=\left\{
\begin{array}{ll}
\Xi           &\text{for} \ \ \ \ d'=1, \\
\Sigma_{L}    &\text{for} \ \ \ \ d'=0,
\end{array}
\right.
\end{equation}
where the universal numbers $\Xi=7.061132...$ and
$\Sigma_{L}=4.317795...$ are the solutions of the scaling
equation~(\ref{a>>1}) for $d'=1$ and $d'=0$, respectively.

In the opposite limit $a\ll1$, for $d'=1$, we get from
Eqs.~(\ref{tl>1}) and~(\ref{2da<1}) the equation
\begin{equation}
\gamma_{E}^{}+\ln\frac{\lambda_{tL}\phi^{1/2}}{4\pi t}=2\sum_{m=1}
^{\infty}K_{0}\left(\frac{\lambda_{tL}}{t}\phi^{1/2}m\right),
\end{equation}
which has
\begin{equation}
\frac{\lambda_{tL}}{t}\chi^{-1/2}=\Xi,
\end{equation}
as an universal solution. For $d'=0$ we have
\begin{equation}\label{last}
\left(\frac{\lambda_{tL}}{t}\phi^{1/2}-6\right)\exp\left(\frac
{\lambda_{tL}}{t}\phi^{1/2}\right)-
\frac{\lambda_{tL}}{t}\phi^{1/2}-6=0
\end{equation}
obtained from Eqs.~(\ref{dp=0}) and~(\ref{2da<1}), where we have used
the identity~(\ref{ident}).

From Eq.~(\ref{last}) we obtain the universal result
\begin{equation}\label{sigmat}
\frac{\lambda_{tL}}{t}\chi^{-1/2}=\Sigma_{t}=6.028966... .
\end{equation}

We finally conclude that if we take $\lambda=\lambda_{c}$ the scaling
functions $Y_{t}^{d'}$ and $Y_{L}^{d'}$ have similar qualitative
behavior weakly depending on the geometry (i.e. bloc $d'=0$ or strip
$d'=1$) of the system. However, for a given geometry one distinguishes
quite different quantitative behavior of the scaling functions
depending on that whether the quantum parameter $\lambda$ is fixed at
its critical value, i.e. $\lambda=\lambda_{c}$, or takes "running"
values $\lambda_{tL}$ obtained from the "shift equations"
~(\ref{lt<1}),~(\ref{tl<1}),~(\ref{tl>1}) or~(\ref{dp=0}).

\subsection{Equation of state}\label{subes}
The equation of state of the model hamiltonian~(\ref{model1}) for
dimensionalities $1<d<3$ is given by (see Eqs.~(\ref{mag})
and~(\ref{finite}))
\begin{eqnarray}
0&=&\delta\lambda- (4\pi)^{-(d+1)/2} \left|
\Gamma \left( \frac{1-d}{2} \right) \right|\left(\frac{h}{\cal M}
\right)^{(d-1)/2} \nonumber\\
& &+\frac{\left(\frac{h}{\cal M}\right)^{(d-1)/2}}{(2\pi)^{(d+1)/2}}
\sum_{m,{\bbox l}(d-d')}^{\hskip7mm\prime} \frac{K_{(d-1)/2}
\left[\left(\frac{h}{\cal M}\right)^{1/2} \left
((\lambda m/t)^{2} + (L |{\bbox l}|)^{2}\right)^{1/2}\right]}
{\left[\left(\frac{h}{\cal M}\right)^{1/2} \left(
(\lambda m/t)^{2} + (L |{\bbox l}|)^{2}\right)^{1/2}\right]
^{(d-1)/2}}+{\cal M}^{2}.\label{statel}
\end{eqnarray}
It is straightforward to write this equation in a similar form as in
Eq.~(\ref{fssl}) or Eq.~(\ref{fsst}), i.e.
\begin{mathletters}\label{stateeq}
\begin{equation}
h={\cal M}^{\delta} f_{h}^{L} \left\{\delta\lambda{\cal M}^{-1/\beta},
\frac{tL}{\lambda},L^{-1/\nu}{\cal M}^{-1/\beta}\right\},
\end{equation}
or
\begin{equation}
h={\cal M}^{\delta} f_{h}^{t} \left\{\delta\lambda{\cal
M}^{-1/\beta},\frac{tL}{\lambda},
\left(\frac{t}{\lambda}\right)^{1/\nu}{\cal M}^{-1/\beta}\right\}.
\end{equation}
\end{mathletters}

Eqs.~(\ref{stateeq}) are generalizations of Eqs.~(\ref{bulk16}) in the
case of systems confined to a finite geometry. The appearance of an
additional variable $\frac{tL}{\lambda}$ is a consequence of the fact
that the system under consideration may be regarded as an
"hyperparallelepiped" (in not necessary an Euclidean space) of linear
size $L$ in $d-d'$ directions and of linear size $L_{\tau}$
in one direction with periodic boundary conditions.

\section{summary}
Since exact solvability is a rare event in statistical physics, the
model under consideration yields a conspicuous possibility to
investigate the interplay of quantum and classical fluctuations as a
function of the dimensionality and the geometry of the system in an
exact manner. Its relation with the QNL$\sigma$M in the
$n\rightarrow\infty$ limit may serve as an illustration of Stanley's
arguments of the relevance of the spherical approximations in the
quantum case. Let us note, however, that the use of such type
arguments needs an additional more subtle treatment in the finite
size case. For this reason we gave a brief overview of the
bulk low--temperature properties, which are similar to those obtained
by saddle--point calculation for the QNL$\sigma$M (see
Section~\ref{bulk}).

The Hamiltonian~(\ref{model1}) can be obtained also in the
"hard--coupling limit" from a more
realistic model that take into account quartic self--interaction
term ${\cal Q}_{\ell}^{4}$, by the anzatz ${\cal Q}_{\ell}^{2}
\Rightarrow\frac{1}{N}\sum_{\ell}{\cal Q}_{\ell}^{2}$ frequently
used in the theory of structural phase transitions (see
Refs.~\onlinecite{verbeure92,pisanova93,chamati94,pisanova95}).
So the Hamiltonian~(\ref{model1}) can be thought as a simple but
rather general model to test some analytical and numerical techniques
in the theory of magnetic and structural phase transitions. The
discussion of the obtained, in Section~\ref{bulk}, results serves
as a basis for the further FSS investigations. Identifying the
temperature, which governs the crossover between the classical and
the quantum fluctuations as an additional temporal dimension one
makes possible the use of the methods of FSS theory in a very
effective way.

In Subsection~\ref{scqp} the shift of the critical quantum parameter
$\lambda$ as a consequence of the quantum and finite--size effects is
obtained. In comparison with the classical case (for details see
Ref.~\onlinecite{chamati96} and references therein) here the problem
is rather complicated by the presence of two finite characteristic
lengths $L$ and $L_{\tau}$. We observe a competition between finite
size and quantum effects which reflects the appearance of two regimes:
low--temperature ($L\gg L_{\tau}$) and very low--temperature ($L\ll
L_{\tau}$). The behavior of the shift is analyzed in some actual cases
of concrete geometries e.g. strip and bloc.

In the parameter space (temperature $t$ and quantum parameter
$\lambda$), where quantum zero point fluctuations are relevant, there
are three distinct regions named "renormalized classical", "quantum
critical", and "quantum disordered". The existence of these regions
in conjunction with both regimes: low--temperature and very
low--temperature make the model a useful tool for the exploration of
the qualitative behavior of an important class of systems.

In Subsection~\ref{zerofs} the susceptibility (or the correlation
length) is calculated and the critical behavior of the system in
different regimes and geometries is analyzed. We have studied the
model~(\ref{model1}) via $\varepsilon$--expansion in order to
illustrate the effects of the dimensionality $d$ on the existence
and properties of the ordered phase. An indicative example is given
by Eqs.~(\ref{cy}) and~(\ref{cy1}), while the former is known (see
Ref.~\onlinecite{Zin89}) the last one is quite different and new.
These shows that one must be accurate in taking the limit $\varepsilon
\rightarrow 0^{+}$ in the context of the FSS theory.

In Subsection~\ref{tdc}, special attention is paid to the
two--dimensional case. The two important cases of strip and bloc
geometries are considered. The universal constant $\Theta$ given by
Eq.~(\ref{theta}), which characterizes the bulk system, is changed to
a set of new universal constants: $\Omega$ (see Eq.~(\ref{Omega})),
$\Xi$ and $\Sigma_{L}$ (see Eq.~(\ref{XiSigma})), and $\Sigma_{t}$
(see Eq.~(\ref{sigmat})). The appearance of new universal constants
reflects the new situation, when there are two relevant values of
the quantum parameter $\lambda$: $\lambda=\lambda_{c}$ in the bulk
case and $\lambda=\lambda_{tL}$ in the case of finite geometries.
Due to their universality these constants may play an important role
even in studying more complicate model Hamiltonians. The behaviors of
the scaling functions at the bulk critical quantum parameter
$\lambda_{c}$ and the shifted critical quantum parameter
$\lambda_{tL}$ are given in FIGs.~\ref{fig2} and~\ref{fig3}.
$L_{\tau}$ is the main characteristic length and the $\frac{1}{L}$
corrections are exponentially small in the case of low--temperature
regime, and vice versa in the case of very low--temperature regime.

The equation of state, for the system confined to the general geometry
$L^{d-d'}\!\!\times\!\!\infty^{d'}\!\!\times\!\!L_{\tau}$, is obtained
in Subsection~\ref{subes}. This reflects the modifications of the
scaling functions as a consequence of the finite sizes and the
temperature.

Finally let us note that this treatment is not restricted to the model
Hamiltonian~(\ref{model1}), but it can be applied to a wide class of
finite lattice models (e.g. directly to the anharmonic crystal model
see Refs.~\onlinecite{verbeure92,pisanova93,chamati94,pisanova95})
and it can also provide a methodology for
seeking different quantum finite--size effects in such systems.

\acknowledgments
The authors thank Dr. Daniel Danchev for valuable discussions on
this work. This work is supported by the Bulgarian Science Foundation
for Scientific Research, Grant No. F-608.

\appendix
\section{}\label{app1}
In this appendix we will derive Eqs.~(\ref{bulk7}) and~(\ref{finite})
of the shifted spherical field $\phi$ for the model
Hamiltonian~(\ref{model1}) confined to the general geometry
$L^{d-d'}\!\!\times\!\!\infty^{d'}\!\!\times\!\!L_{\tau}$,
with periodic boundary conditions, in the low temperature regime.
To achieve that, let us start with Eq.~(\ref{model3})
\begin{mathletters}\label{self3}
\begin{equation}\label{a1}
1= {\cal W}_{d}(\phi,L,t)  + \frac{h^{2}}{\phi^{2}},
\end{equation}
where we have used the notation
\begin{equation}\label{W(f)}
{\cal W}_{d}(\phi,L,t)=\frac{t}{N}\sum_{m=-\infty}^{\infty}
\sum_{\bbox q} \frac{1}{\phi+2\sum_{i=1}^{d}(1-\cos q_{i})+
\left(\frac{2\pi t}{\lambda}\right)^{2} m^{2}}.
\end{equation}
\end{mathletters}

Now if we assume that the system is infinite in $d'$ dimensions, then we
may write Eq.~(\ref{W(f)}) in the following form
\begin{eqnarray}
{\cal W}_{d}(\phi,L,t)&=&\frac{t L^{d'-d}}{(2\pi)^{d'}}
\sum_{{\bbox q}(d-d')} \sum_{m=-\infty}^{\infty} \nonumber\\
& &\int_{-\pi}^{\pi} d^{d'}{\bbox q} \int_{0}^{\infty} dx
\exp\left\{ -x\left(\phi+2\sum_{i}(1-\cos q_{i})
+\left(\frac{2\pi t}{\lambda}\right)^{2}m^{2}\right)\right\},
\label{fg1}
\end{eqnarray}

To obtain the last expression use has been made of the representation
\begin{equation}\label{IR}
\frac{1}{z}=\int_{0}^{\infty} \exp(-zx) dx
\end{equation}
and that $N=L^{d}$.

Now by rearranging it is possible to write Eq.~(\ref{fg1})
in the following form
\begin{eqnarray}
{\cal W}_{d}(\phi,L,t)&=& t \sum_{m=\infty}^{\infty} \int_{0}^{\infty}
dx \exp\left[-x\left\{\phi+2d+\left(\frac{2\pi t}{\lambda}\right)^{2}
m^{2}\right\}\right] \left[I_{0}(2x)\right]^{d'} \nonumber \\
& &\times \left[\frac{1}{L}\sum_{q} \exp(2x\cos
q)\right]^{d-d'}\label{fg2}.
\end{eqnarray}
Here $I_{0}(x)$ is the modified Bessel function.

The use the of Poisson summation formula
\begin{equation}\label{poisson}
\frac{1}{L}\sum_{n=-\frac{L-1}{2}}^{n=\frac{L-1}{2}}G\left(\frac{2\pi
n}{L}\right)
= \sum_{l=-\infty}^{\infty} \int_{-\pi}^{\pi} \frac{dq}{2\pi} G(q)
\exp(iqlL),
\end{equation}
where $G(q)$ is a periodic function,
allows us to continue the sum over the wave vector $q=2\pi n/L$ ($n\in
[-L/2,L/2]$) to the rest of the real line periodically. With the aid
of Eq.~(\ref{poisson}) we can transform (\ref{fg2}) into
\begin{equation}\label{fg3}
{\cal W}_{d}(\phi,L,t)= t \sum_{m=\infty}^{\infty} \int_{0}^{\infty}
dx \exp\left[-x\left\{\phi+2d+\left(\frac{2\pi
t}{\lambda}\right)^{2}m^{2}\right\}\right] \left[I_{0}(2x)\right]^{d'}
\left[\sum_{l=-\infty}^{\infty} I_{lL} (2x)\right]^{d-d'}.
\end{equation}

In order to investigate the low--temperature effects for the model
Hamiltonian~(\ref{model1}) we use the Jacobi identity
\begin{equation}\label{jacobi}
\sum_{m=-\infty}^{\infty} \exp\left(-um^{2}\right)=
\left(\frac{\pi}{u}\right)^{1/2}\sum_{m=-\infty}^{\infty}
\exp\left(-m^{2}\frac{\pi^{2}}{u}\right),
\end{equation}
which applied to Eq.~(\ref{fg3}) gives
\begin{eqnarray}
{\cal W}_{d}(\phi,L,t)&=&\lambda {\cal W}_{d}(\phi)+\frac{\lambda}{2
\pi^{1/2}}\nonumber \\
& &\times\sum_{m,{\bbox l}(d-d')}^{\hskip7mm\prime}\int_{0}^{\infty}
\frac{dx}{x^{1/2}}\exp\left[-x(\phi+2d)-\left(\frac{\lambda}{2tx}
\right)^{2}m^{2}\right]\left[I_{0}(2x)\right]^{d'}I_{{\bbox l}L}(2x)
\label{fg4},
\end{eqnarray}
where we have used the formal notations
$$
\sum_{{\bbox l}(d-d')}I_{{\bbox l}L}(2x)=
\left[\sum_{l}I_{lL}(2x)\right]^{d-d'},
\ \ {\bbox l}^{2} = l_{1}^{2}+\cdots+l_{d-d'}^{2},
$$
\begin{equation}\label{aa}
{\cal W}_{d} (\phi)=\frac{1}{2(2\pi)^{d}}\int_{-\pi}^{\pi}dq_{1}
\cdots\int_{-\pi}^{\pi}dq_{d}
\left(\phi+2\sum_{i=1}^{d} (1-\cos q_{i})\right)^{-1/2}.
\end{equation}
the prime means that the vector with components
$m=l_{1}=\cdots=l_{d-d'}=0$ is omitted.

At sufficiently low--temperature ($\frac{\lambda}{t}\gg1$) and large
enough size ($L\gg1$) we can use the asymptotic form for the Bessel
functions~\cite{singh85}
\begin{equation}
I_{\nu}(x) \approx \frac{e^{x-\nu^2/2x}}{\sqrt{2\pi x}}\left[1+
\frac{1}{8x}+\frac{9-32\nu^{2}}{2!(8x)^{2}}+\cdots\right],
\end{equation}
in order to get after substitution in Eq.~(\ref{a1})
\begin{equation}\label{rev1}
1=\lambda {\cal W}_{d}(\phi) +
\frac{\lambda\phi^{(d-1)/2}}{(4\pi)^{(d+1)/2}}\sum_{m,{\bbox l}(d-d')}
^{\hskip7mm\prime}\frac{K_{\frac{d-1}{2}}\left[\phi^{1/2}\left(\left(
\frac{\lambda}{t}\right)^{2}m^2+L^{2} {\bbox l}^{2} \right)^{1/2}
\right]}{\left[\phi^{1/2}\left(\left(\frac{\lambda}{t}\right)^{2}
m^2+L^{2}{\bbox l}^{2}\right)^{1/2}\right]^{\frac{d-1}{2}}}.
\end{equation}

The Watson type integral ${\cal W}_{d} (\phi)$ (see Eq.~(\ref{aa}))
has been studied in
considerable details;\cite{pisanova93} for $1<d<3$,
it can be approximated by ($\phi \ll 1$),
\begin{equation}\label{bulk5}
{\cal W}_{d}(\phi)\approx{\cal W}_{d}(0)-(4\pi)^{-(d+1)/2}\left|
\Gamma \left( \frac{1-d}{2} \right) \right|\phi^{(d-1)/2},
\end{equation}
which leads one to conclude that at zero temperature the system
exhibits a phase transition driven by the parameter $\lambda$ at
the quantum critical point:
\begin{equation}\label{bulk6}
\lambda_{c}=\frac{1}{{\cal W}_{d}(0)}.
\end{equation}
Finally,
substituting~(\ref{bulk5}) in~(\ref{rev1}) we obtain Eq.~(\ref{finite}).
Eq.~(\ref{bulk7}) is obtained by setting $d=d'$.

\section{}\label{app2}
In this appendix we will sketch a way to find the asymptotic behavior
of the functions ${\cal K}_{a} (\nu|p,y)$ defined in
section~\ref{fing} (see Eq.~(\ref{fssl})). They have the following
form
\begin{mathletters}\label{b1}
\begin{equation}
{\cal K}_{a} (\nu|p,y) = \sum_{m,{\bbox l}(p-1)}^{\hskip7mm\prime}
\frac{K_{\nu}\left(2y\sqrt{{\bbox l}^{2}+a^{2}m^{2}}\right)}
{\left(y\sqrt{{\bbox l}^{2}+a^{2}m^{2}}\right)^{\nu}}, \ \ y>0,
\end{equation}
where
\begin{equation}\label{bfq}
{\bbox l}^2 = l_{1}^{2}+l_{2}^{2}+\cdots+l_{p-1}^{2}.
\end{equation}
\end{mathletters}

By the use of the integral representation of the modified Bessel
function
\begin{equation}
K_{\nu}\left(2\sqrt{zt}\right)=K_{-\nu}\left(2\sqrt{zt}\right)=
\frac{1}{2}\left(\frac{z}{t}\right)^{\nu/2}
\int_{0}^{\infty}x^{-\nu-1}e^{-tx-z/x}dx
\end{equation}
and the Jacobi identity for a $p$--dimensional lattice sum
\begin{equation}\label{djacobi}
\sum_{m,{\bbox l}(p-1)} e^{-({\bbox l}^{2}+a^{2}m^{2})t}= \frac{1}{a}
\left(\frac{\pi}{t}\right)^{p/2}
\sum_{m,{\bbox l}(p-1)} e^{-\pi^{2} ({\bbox l}^{2}+m^{2}/a^{2})/t},
\end{equation}
we may write (\ref{b1}) as
\begin{eqnarray}
{\cal K}_{a} (\nu|p,y)&=&\frac{\pi^{p/2}}{2a} \Gamma\left(\frac{p}{2}
-\nu\right) y^{-p}  \nonumber \\
& &+\frac{\pi^{2\nu-p/2}}{2a} y^{-2\nu} \int_{0}^{\infty} dx
x^{\frac{1}{2}p-\nu-1} e^{-x y^{2}/\pi^{2}}
\left[\sum_{m,{\bbox l}(p-1)}^{\hskip7mm\prime} e^{-x({\bbox l}^{2}+
m^{2}/a^{2})}-a \left(\frac{\pi}{x}\right)^{p/2}\right].\label{A2}
\end{eqnarray}
Let us notice that the two terms in the square brackets in the last
equality cannot be integrated separately, since they diverge.
Nevertheless, in order to encounter this divergence,
we can transform further (\ref{A2}) by adding and
subtracting the unity from $\exp(-x y^{2}/\pi^{2})$, which
enables us to write down (after some algebra) the result
\begin{eqnarray}
{\cal K}_{a} (\nu|p,y)&=&\frac{\pi^{p/2}}{2a} \Gamma\left(\frac{p}{2}
-\nu\right)y^{-p}+\frac{\pi^{2\nu-p/2}}{2a} y^{-2\nu} C_{a}(p|\nu)
-\frac{1}{2} \Gamma(-\nu) \nonumber \\
& &+\frac{\pi^{2\nu-p/2}}{2a} \frac{\Gamma\left(\frac{p}{2}-\nu\right)}
{y^{2\nu}}\sum_{m,{\bbox l}(p-1)}^{\hskip7mm\prime}
\left[\left({\bbox l}^{2} +\frac{m^{2}}{a^{2}}+\frac{y^{2}}{\pi^{2}}
\right)^{\nu-p/2}-\left({\bbox l}^{2}+\frac{m^{2}}{a^{2}}\right)
^{\nu-p/2}\right],\label{k(n,m)}
\end{eqnarray}
where
\begin{mathletters}\label{madelung}
\begin{eqnarray}
C_{a}(p|\nu) &=& \lim_{\delta\rightarrow0}\int_{\delta}^{\infty} dx
x^{\frac{1}{2}p-\nu-1} \left[\sum_{m,{\bbox l}(p-1)}^{\hskip7mm\prime}
e^{-x({\bbox l}^{2}+m^{2}/a^{2})}-a \left(\frac{\pi}{x}\right)
^{p/2}\right] ,\\
&=&\lim_{\delta\rightarrow0}\left\{\sum_{m,{\bbox l}(p-1)}
^{\hskip7mm\prime}\frac{\Gamma\left[\frac{p}{2}-\nu,\delta
\left({\bbox l}^{2}+\frac{m^{2}}{a^{2}}\right)\right]}{\left(
{\bbox l}^{2}+\frac{m^{2}}{a^{2}}\right)^{\nu-p/2}}\right.\nonumber\\
& & \ \ \ \ \ \ \ \ \ \ \ \ \ \ \ -
\left.\int_{-\infty}^{\infty}\cdots\int_{-\infty}^{\infty}dmd^{p-1}
{\bbox l}\frac{\Gamma\left[\frac{p}{2}-\nu,\delta\left({\bbox l}^{2}
+\frac{m^{2}}{a^{2}}\right)\right]}
{\left({\bbox l}^{2}+\frac{m^{2}}{a^{2}}\right)^{\nu-p/2}}\right\}
\end{eqnarray}
\end{mathletters}
is the Madelung--type constant and $\Gamma[\alpha,x]$ is the
incomplete gamma function.

We see from Eq.~(\ref{k(n,m)}) that the shift of the critical quantum
parameter is given by the Madelung type constant~(\ref{madelung})
instead of the sum in Eq.~(\ref{shift}). Indeed it is possible to show
that these two representations are equivalent. This may be done,
following Ref.~\onlinecite{singh89}, by starting from the Jacobi
identity Eq.~(\ref{djacobi}), where we multiply the two sides by
$\delta^{p/2-\nu-1}$ and integrating over $\delta$ to obtain the key
equation
\begin{eqnarray}C_{a}(p|\nu)&=&\sum_{m,{\bbox l}(p-1)}^{\hskip7mm\prime}
\frac{\Gamma\left[\frac{p}{2}-\nu,\delta\left({\bbox l}^{2}+\frac{m^{2}}
{a^{2}}\right)\right]}{\left({\bbox l}^{2}+\frac{m^{2}}{a^{2}}\right)
^{p/2-\nu}}-\frac{\delta^{p/2-\nu}}{\frac{p}{2}-\nu}\nonumber\\
& &\ \ \ +a\pi^{p/2-2\nu}\sum_{m,{\bbox l}(p-1)}^{\hskip7mm\prime}
\frac{\Gamma\left[\nu,\frac{\pi^{2}\left({\bbox l}^{2}+ a^{2}m^{2}
\right)}{\delta}\right]}{\left({\bbox l}^{2}+ a^{2}m^{2}\right)
^{\nu}}-a\frac{\pi^{p/2}}{\nu \delta^{\nu}}.\label{sp}
\end{eqnarray}

Finally form Eq.~(\ref{sp}) we see easily that the integration
constant $C_{a}(p|\nu)$ may be written in two different forms. In the
first case we take the limit $\delta\rightarrow\infty$ and obtain
\begin{equation}
C_{a}(p|\nu) = a\pi^{p/2-2\nu}\Gamma(\nu)
\sum_{m,{\bbox l}(p-1)}^{\hskip7mm\prime} \frac{1}{\left({\bbox l}^{2}+
a^{2}m^{2}\right)^{\nu}}.
\end{equation}
In the other case we take the limit $\delta\rightarrow0$, and then both
first and last terms in the rhs of Eq.~(\ref{sp}) yields
Eq.~(\ref{madelung}).

Using a similar procedure we find, for the functions
$\tilde{\cal K}_{a}(\nu|p,y)$ defined in~(\ref{katt1}), the following
expression
\begin{eqnarray}
\tilde{\cal K}_{a} (\nu|p,y)&=&\frac{\pi^{p/2}}{2a^{p-1}} \Gamma\left(
\frac{p}{2}-\nu\right)y^{-p}
+\frac{\pi^{2\nu-p/2}}{2a^{p-1}} y^{-2\nu} \tilde{C}_{a}(p|\nu)
-\frac{1}{2} \Gamma(-\nu) \nonumber \\
& &+\frac{\pi^{2\nu-p/2}}{2a^{p-1}} \frac{\Gamma\left(\frac{p}{2}-\nu
\right)}{y^{2\nu}}\sum_{m,{\bbox l}(p-1)}^{\hskip7mm\prime}
\left[\left(\frac{{\bbox l}^{2}}{a^{2}} +m^{2}+\frac{y^{2}}
{\pi^{2}}\right)^{\nu-p/2}-\left(\frac{{\bbox l}^{2}}{a^{2}}
+m^{2}\right)^{\nu-p/2}\right].\label{lastap}
\end{eqnarray}
Here the Madelung type constant is given by
\begin{mathletters}\label{madelungt}
\begin{eqnarray}
\tilde{C}_{a}(p|\nu)&=&\lim_{\delta\rightarrow0}\int_{\delta}^{\infty}
dx x^{\frac{1}{2}p-\nu-1}\left[\sum_{m,{\bbox l}(p-1)}^{\hskip7mm\prime}
e^{-x({\bbox l}^{2}/a^{2}+m^{2})}-
a^{p-1} \left(\frac{\pi}{x}\right)^{p/2}\right] ,\\
&=&\lim_{\delta\rightarrow0}\left\{\sum_{m,{\bbox l}(p-1)}
^{\hskip7mm\prime}\frac{\Gamma\left[\frac{p}{2}-\nu,\delta\left(
\frac{{\bbox l}^{2}}{a^{2}}+m^{2}\right)\right]}{\left(\frac{{\bbox l}
^{2}}{a^{2}}+m^{2}\right)^{\nu-p/2}} \right.\nonumber \\
& & \ \ \ \ \ \ \ \ \ \ \ \ \ \ \ -
\left.\int_{-\infty}^{\infty}\cdots\int_{-\infty}^{\infty}dmd^{p-1}
{\bbox l}\frac{\Gamma\left[\frac{p}{2}-\nu,\delta\left(\frac{{\bbox l}
^{2}}{a^{2}}+m^{2}\right)\right]}{\left(\frac{{\bbox l}^{2}}{a^{2}}
+m^{2}\right)^{\nu-p/2}}\right\}\\
&=&a^{p-1}\pi^{p/2-2\nu}\Gamma(\nu)\sum_{m,{\bbox l}(p-1)}
^{\hskip7mm\prime} \frac{1}{\left(a^{2}{\bbox l}^{2}+ m^{2}\right)^{\nu}}.
\end{eqnarray}
\end{mathletters}

Eqs.~(\ref{k(n,m)}) and~(\ref{lastap}) are slight generalizations (for
the anisotropic case $a\neq1$) of the result obtained in
Ref.~\onlinecite{singh89} from one side, and get in touch with the
Watson type sums proposed earlier in Ref.~\onlinecite{brankov88} from
the other (see also Ref.~\onlinecite{chamati96}).

If we put in Eqs.~(\ref{k(n,m)}) or~(\ref{lastap}) $d=2$, $d'=0$ and
$a=1$  we obtain the identity
\begin{eqnarray}
\sum_{l_{1},l_{2}}^{\hskip7mm\prime}\frac{\exp\left(-y
\sqrt{l_{1}^{2}+l_{2}^{2}}\right)}{\sqrt{l_{1}^{2}+l_{2}^{2}}}&=&
\frac{2\pi}{y}+4\zeta\left(\frac{1}{2}\right)
\beta\left(\frac{1}{2}\right)+y \nonumber\\
& &+2\pi\sum_{l_{1},l_{2}}
^{\hskip7mm\prime}\left\{\frac{1}{\sqrt{y+4\pi^{2}(l_{1}^{2}+
l_{2}^{2})}}-\frac{1}{2\pi\sqrt{l_{1}^{2}+l_{2}^{2}}}\right\}.
\label{ident}
\end{eqnarray}


\begin{references}
\bibitem{dyn} In this approximation the dynamical critical exponent
$z$ is equal to the unity i.e. there is no  anisotropy between the
space and time dimensions.
\end{references}
\end{document}